\documentclass[onecolumn,superscriptaddress,showpacs,preprintnumbers,pra]{revtex4}
\usepackage{epsfig}
\usepackage{amssymb}
\usepackage{amsmath}
\usepackage{epstopdf}
\usepackage{graphicx,color}  
\usepackage{bm}  
\usepackage{epstopdf}

\begin{document}

\title{Single-Particle Momentum Distributions of Efimov States in Mixed-Species Systems}
\author{M.~T. Yamashita}
\affiliation{Instituto de F\'\i sica Te\'orica, UNESP - Univ Estadual Paulista, C.P. 70532-2, CEP 01156-970, S\~ao Paulo, SP, Brazil}
\author{F.~F. Bellotti}
\affiliation{Instituto Tecnol\'{o}gico de Aeron\'autica, 12228-900, S\~ao Jos\'e dos Campos, SP, Brazil}
\affiliation{Department of Physics and Astronomy, Aarhus University, DK-8000 Aarhus C, Denmark}
\affiliation{Instituto de Fomento e Coordena\c{c}{\~a}o Industrial, 12228-901, S{\~a}o Jos{\'e} dos Campos, SP, Brazil}
\author{T. Frederico}
\affiliation{Instituto Tecnol\'{o}gico de Aeron\'autica, 12228-900, S\~ao Jos\'e dos Campos, SP, Brazil}
\author{D.~V. Fedorov}
\author{A.~S. Jensen}
\author{N.~T. Zinner}
\affiliation{Department of Physics and Astronomy, Aarhus University, DK-8000 Aarhus C, Denmark}
\date{\today }

\begin{abstract}
We solve the three-body bound state problem in three dimensions 
for mass imbalanced systems of two identical bosons
and a third particle in the universal limit where the interactions
are assumed to be of zero-range. The system displays the Efimov effect and 
we use the momentum-space wave equation to derive formulas for the scaling 
factor of the Efimov spectrum for any mass ratio assuming either that 
two or three of the two-body subsystems have a bound state at zero energy.
We consider the single-particle momentum distribution analytically and 
numerically and analyse the tail of the momentum distribution to obtain 
the three-body contact parameter. 
Our finding demonstrate that the functional form of the 
three-body contact term depends on the mass ratio and we obtain an 
analytic expression for this behavior.
To exemplify our results, we consider mixtures of Lithium with either
two Caesium or Rubium atoms which are systems of current experimental 
interest.
\end{abstract}

\maketitle

\section{Introduction}
Few- and many-body systems in the presence of strong interactions is an
increasingly fruitful direction in physics due to several advances in
the experimental realization of such systems within cold atomic gases
\cite{bloch2008}. One aspect of this pursuit concerns the so-called
unitary regime in which the two-body scattering amplitude saturates
the unitarity bound of basic quantum mechanics. For neutral and
non-polar cold atomic gases interactions are very short-ranged
and one can often use the universal limit where interactions
are modelled by a zero-range potential which is then subsequently
connected to the low-energy two-body scattering dynamics
through the scattering length, $a$. In this framework, the
unitary regime is simply characterized by a scattering length
that is much larger than any other relevant scale in the
system under study. The problem of characterizing a
strongly interacting system at unitarity in turn lacks a
natural scale for the interactions, and one expects
universal behavior of structural and dynamical properties
that should be applicable irrespective of the details
of the short-distance physics, and thus have applicability
in many different subfields.

An important recent development was the derivation of a
number of universal relations that describe the physics of
unitary two-component Fermi gases by Shina Tan \cite{tan2008}. It
turns out that a parameter dubbed the (two-body) contact, $C_2$,
emerges in expressions for both structural (energy, adiabatic theorem,
pressure, virial theorem \cite{werner2012a}) and dynamical observables (inelastic
loss \cite{braaten2008}, radio-frequency spectroscopy \cite{braaten2010,langmack2012}).
This provides a strong connection between the few-body quantities
and many-body observables, a connection that has by now been
experimentally verified \cite{stewart2010,kuhnle2010}. The
relations are not particular to the unitary Fermi gas
but may also be applied to bosons
\cite{combescot2009,schakel2010,braaten2011,castin2011,valiente2011,werner2012b}
as another recent experimental effort has confirmed \cite{wild2012}.

Another direction that has enjoyed access to the unitary regime
is the study of universal three-body bound states and the famous
Efimov effect \cite{efi70}. The effect occurs close to the
unitary point and implies that a sequence of three-body bound states
occurs wherein two successive states always have the same fixed
ratio of their binding energies.
The effect was first observed in cold atomic gas
experiments \cite{kraemer2006} and has subsequently opened a
new research direction dubbed Efimov physics \cite{ferlaino2010}.
An interesting recent finding is that the presence of the Efimov
effect implies an extension of the universal relations discussed
above, and the introduction of a three-body contact parameter, $C_3$
\cite{braaten2011,castin2011}. This parameter vanishes in the
case of two-component Fermi gases since the Pauli principle
suppresses three-body correlations at short distances \cite{werner2012a}.

In this paper we study three-body bound Efimov-like states
for systems that contain two identical bosons
and a third distinguishable particle. Our goal is to address
the contact parameters of such systems when the masses are
different and for different strengths of the interaction
parameters by computing the single-particle
momentum distributions and studying its asymptotic behavior.
We will consider only the universal regime, i.e.
we approximate all two-body potentials by zero-range interactions.
In comparison to the previous studies containing three identical
bosons, we find that for different particles there are additional
contributions to the asymptotic behavior of the momentum
distributions. This suggests that such measurements are a
useful probe to descriminate between effects of identical
and non-identical three-body correlations in the many
current cold gas experiments that have mixtures of different
kinds of atoms in the same trap.

The paper is organized as follows. In Sec.~\ref{formal} we
introduce the momentum-space formalism that we use and
Sec.~\ref{asymp}
contains a discussion of the three-body wave function (or spectator function).
In Sec.~\ref{asymp} we also present an analytic
derivation of the Efimov scaling factors as
function of mass ratio in the cases where either two or three of the 
two-body subsystems have a two-body bound state at zero energy.
We proceed to present our results for the asymptotic momentum
distributions in Sec.~\ref{amomentum} and demonstrate that the 
subleading correction has a functional form that depends on the 
mass ratio. The details of the analytic derivations of the 
asymptotic momentum behavior are given in Appendix~\ref{appendix1}
for completeness.
Results
for relevant experimental mixtures of
$^6$Li-$^{133}$Cs-$^{133}$Cs and $^6$Li-$^{87}$Rb-$^{87}$Rb
are discuss in Sec.\ref{results} for different choices of the
interaction parameters. Sec.~\ref{diss} contains conclusions
and outlook for future studies.

\section{Formalism}\label{formal}
We consider a system that has an $AAB$ structure, where the two
$A$ particles are identical bosons and the third $B$ particle is
of a different kind.
When we discuss our results below, we will
focus on two combinations that are of interest
to current experimental efforts in cold atoms, $A=^{133}$Cs, $B=^{6}$Li
and $A=^{87}$Rb, $B=^{6}$Li. 

Since we are interested in the universal limit where the range
of the two-body potentials can be neglected we consider purely
zero-range interactions in the following. More precisely, if
$r_0$ is the range of the two-body potential, we are assuming that
the scattering length, $a$, is $a\gg r_0$. For simplicity we will
use units where $\hbar=m_A=1$ from now on.
After partial wave
projection, the $s-$wave coupled subtracted integral equations for the
spectator functions, $\chi$, and the absolute value of the 
three-body binding energy, $E_3$ are given by \cite{amorim1997,bel11,frederico2012} 
\begin{eqnarray}
\chi_{A A}(y)&=&2\tau_{A A}(y;E_3) \int_0^\infty dx
\frac{x}{y} G_1(y,x;E_3)\chi_{A B}(x)
\label{chi1} \\
\chi_{A B}(y)&=&\tau_{A B}(y;E_3) \int_0^\infty dx
\frac{x}{y} \left[G_1(x,y;E_3) \chi_{A A}(x) + {\cal A}
G_2(y,x;E_3) \chi_{A B}(x)\right] ; \label{chi2}
\end{eqnarray}
\begin{eqnarray}
\tau_{A A}(y;E_3)&\equiv &\frac{1}{\pi}
\left[\sqrt{E_3+\frac{{\cal A}+2}{4{\cal A}} y^2} \mp \sqrt{E_{AA}}
\right]^{-1},  \label{tau1}
\\
\tau_{A B}(y;E_3)&\equiv
&\frac{1}{\pi}\left(\frac{{\cal A}+1}{2{\cal A}}\right)^{3/2}
\left[\sqrt{E_3+\frac{{\cal A}+2}{2({\cal A}+1)} y^2} \mp
\sqrt{E_{AB}}\right]^{-1}\ , \label{tau2}
\\
G_1(y,x;E_3)&\equiv &\log
\frac{2{\cal A}(E_3+x^2+xy)+y^2({\cal A}+1)}{2{\cal A}(E_3+x^2-xy)+y^2({\cal A}+1)}
-\log\frac{2{\cal A}(\mu^2+x^2+xy)+y^2({\cal A}+1)}{2{\cal A}(\mu^2+x^2-xy)+y^2({\cal A}+1)} ,
\label{G1} \\
G_2(y,x;E_3)&\equiv & \log \frac{2({\cal A}E_3
+xy)+(y^2+x^2)({\cal A}+1)}{2({\cal A} E_3-xy)+(y^2+x^2)({\cal A}+1)}
-\log \frac{2({\cal A}\mu^2 +xy)+(y^2+x^2)({\cal A}+1)}{2({\cal A}\mu^2 -xy)+(y^2+x^2)({\cal A}+1)},
\label{G2}
\end{eqnarray}
where $x$ and $y$ denote (dimensionless) momenta.
Here we have introduced the mass number ${\cal A}=m_B/m_A$. The interaction energies
of the $AA$ and $AB$ subsystems are parametrized by $E_{AA}$ and $E_{AB}$,
and the plus and minus
signs in \eqref{tau1} and \eqref{tau2} refer to virtual and bound two-body subsystems,
respectively \cite{yamashita2002,bringas2004,yamashita2008}.
We map $E_{AA}$ and $E_{AB}$ into the usual scattering lengths,
$a_{AA}$ and $a_{AB}$ through the relation $E\propto |a|^{-2}$.
This relation typically
holds for broad resonances and a more detailed mapping needs to be done in the
general case \cite{chin2010}. Throughout most of this work we will focus on the
region close to unitarity in the $AB$ system, i.e. $|a_{AB}|\to \infty$ or $E_{AB}\to 0$.
In light of the fact that experimental information about mixed systems of the
$AAB$ type is still sparse, we will consider the two extreme cases i) $E_{AA}=0$
and ii) a non-interacting $AA$ subsystem.

In the numerical work presented later on we will set $\mu^2=1$ for the subtraction point (see for
instance Ref.~\cite{frederico2012} for a detailed discussion and references).
On the other hand, in the analytical derivations we will take the limit $\mu\to\infty$.
We note that this subtraction method is basically equivalent to the procedure
employed by Danilov \cite{danilov1961} to regularize the original three-body
Skorniakov-Ter-Martirosian equation \cite{stm1956}. A very detailed recent
discussion of these issues was given by Pricoupenko \cite{pri2010,pri2011}.

Defining as $\vec{k}_\alpha \,\,\, (\alpha=i,j,k)$ the momenta of each particle in 
the rest frame, we have that the Jacobi momenta from one particle to the center-of-mass 
of the other two and the relative momentum of the two are given, respectively, by
\begin{equation}
\vec{q}_k=m_{ij,k}\left(\frac{\vec{k}_k}{m_k}-\frac{\vec{k}_i+\vec{k}_j}{m_i+m_j}\right)=
\vec{k}_k \,\,\,\,{\rm and}\,\,\,\,\, \vec{p}_k=m_{ij}\left(\frac{\vec{k}_i}{m_i}-
\frac{\vec{k}_j}{m_j}\right),
\end{equation}
where $\{i,j,k\}$ is an even permutation of the particles $\{A,A^\prime,B\}$ and
we have used that $\vec{k}_i+\vec{k}_j+\vec{k}_k=0$ in the center of mass system.
The reduced masses are defined such that $m_{ij}=\frac{m_im_j}{m_i+m_j}$ and 
$m_{ij,k}=\frac{m_k(m_i+m_j)}{m_i+m_j+m_k}$.

Below we define exactly what we
mean by single-particle momentum distributions for particles
of type $A$ and type $B$.
For a zero-range potential the three-body wave
function for an $AAB$ system, composed by two identical particles
$A$ and one different $B$, can be written in terms of the spectator
functions in the basis $|\vec{q}_B\vec{p}_B\rangle$ as
\begin{eqnarray}
\langle\vec{q}_B\vec{p}_B|\Psi\rangle&=&
\frac{\chi_{AA}(q_i)+\chi_{AB}(q_j)+\chi_{AB}(q_k)}{E_3+H_0}
=\frac{\chi_{AA}(q_B)+\chi_{AB}(|\vec{p}_B-\frac{\vec{q}_B}{2}|)
+\chi_{AB}(|\vec{p}_B+\frac{\vec{q}_B}{2}|)}{E_3+H_0},
\label{psiqb}
\end{eqnarray}
or in the basis $|\vec{q}_A\vec{p}_A\rangle$ as
\begin{eqnarray}
\langle\vec{q}_A\vec{p}_A|\Psi\rangle&=&
\frac{\chi_{AA}(|\vec{p}_A-\frac{{\cal A}}{{\cal A}+1}\vec{q}_A|)+
\chi_{AB}(|\vec{p}_A+\frac{1}{{\cal A}+1}\vec{q}_A|)+\chi_{AB}(q_A)}
{E_3+H_0^\prime},
\end{eqnarray}
where $H_0=\frac{p_B^2}{2m_{AA}}+\frac{q_B^2}{2m_{AA,B}}$ and
$H_0^\prime=\frac{p_A^2}{2m_{AB}}+\frac{q_A^2}{2m_{AB,A}}$. The reduced masses
are given by $m_{AA}=\frac12$, $m_{AA,B}=\frac{2{\cal A}}{{\cal A}+2}$,
$m_{AB}=\frac{{\cal A}}{{\cal A}+1}$ and $m_{AB,A}=\frac{{\cal A}+1}{{\cal A}+2}$.

The momentum distributions for the particles $A$ and $B$ are
\begin{equation}
\label{nqaqb}
n(q_B)=\int d^3p_B |\langle\vec{q}_B\vec{p}_B|\Psi\rangle|^2,
\hspace{1cm} n(q_A)=\int d^3p_A |\langle\vec{q}_A\vec{p}_A|\Psi\rangle|^2
\end{equation}
and they are normalized such that $\int d^3q n(q)=1$. Note that our definition
of momentum distributions as well as their normalizations differ from Ref.~\cite{castin2011}. 
In Ref.~\cite{castin2011} there is a factor of $1/(2\pi)^3$ multiplying the definition of
$n(q)$, which is normalized to 3, the number of particles.

\section{Asymptotic formulas for the spectator functions}\label{asymp}
We now consider the asymptotic behavior of the spectator function to
derive some analytic formulas and compare to corresponding numerical
results. To access the large momentum regime $\sqrt{E_3}\ll
q$, we take the limit $\mu\to\infty$ and
$E_3=E_{AA}=E_{AB}\to 0$. The coupled equations
for the spectator functions consequently simplify and become
\begin{eqnarray}
\chi_{A A}(y)&=&\frac{2}{\pi} \left[y\sqrt{\frac{{\cal A}+2}{4{\cal A}}}
\right]^{-1} \int_0^\infty dx \frac{x}{y} G_{1a}(y,x)\chi_{A B}(x)
\label{chi1a} \\
\chi_{A B}(y)&=&
\frac{1}{\pi}\left(\frac{{\cal A}+1}{2{\cal A}}\right)^{3/2}
\left[y\sqrt{\frac{{\cal A}+2}{2({\cal A}+1)}}\right]^{-1}
 \int_0^\infty dx \frac{x}{y}
\left[G_{1a}(x,y) \chi_{A A}(x) + {\cal A} G_{2a}(y,x) \chi_{A B}(x)\right]
; \label{chi2a}
\end{eqnarray}
where
\begin{eqnarray}
G_{1a}(y,x)&\equiv &\log
\frac{2{\cal A}(x^2+xy)+y^2({\cal A}+1)}{2{\cal A}(x^2-xy)+y^2({\cal A}+1)}
\label{G1a} \\
G_{2a}(y,x)&\equiv & \log \frac{
(y^2+x^2)({\cal A}+1)+2xy}{(y^2+x^2)({\cal A}+1)-2xy} \label{G2a}
\end{eqnarray}

We now proceed to solve these equations by using the ans{\"a}tze
\begin{equation}
\chi_{A A}(y)=c_{AA}\; y^{-2+\imath s} ~~~{\text{ and }} ~~~~ \chi_{A
B}(y)=c_{AB}\;y^{-2+\imath s}, \label{sol}
\end{equation}
where $y$ once again denotes a (dimensionless) momentum.
Inserting the functions \eqref{sol} in the set of coupled
equations and performing the scale transformation $x=y\;z$, in the
integrand of Eqs.~\eqref{chi1a} and \eqref{chi2a}, one has the
following set of equations
\begin{eqnarray}
c_{AA}&=&c_{AB}\frac{2}{\pi}\sqrt{\frac {4{\cal A}}{{\cal A}+2}} \int_0^\infty
dz\; z^{-2+1+\imath s} \; \log
\frac{2{\cal A}(z^2+z)+({\cal A}+1)}{2{\cal A}(z^2-z)+({\cal A}+1)}
\label{chi1a1} \\
c_{A B}&=&
\frac{1}{\pi}\left(\frac{{\cal A}+1}{2{\cal A}}\right)^{3/2} \sqrt{\frac
{2({\cal A}+1)}{{\cal A}+2}}
 \int_0^\infty dz\; z^{-2+1+\imath s}
\left[c_{A A}\;\log
\frac{2{\cal A}(1+z)+z^2({\cal A}+1)}{2{\cal A}(1-z)+z^2({\cal A}+1)} \right. \nonumber \\
&&+\left. {\cal A} \;
 c_{A B}\;
\log \frac{(1+z^2)({\cal A}+1)+2z}{(1+z^2)({\cal A}+1)-2z}
 \right] . \label{chi2a1}
\end{eqnarray}

\begin{figure}[htb!]
\epsfig{file=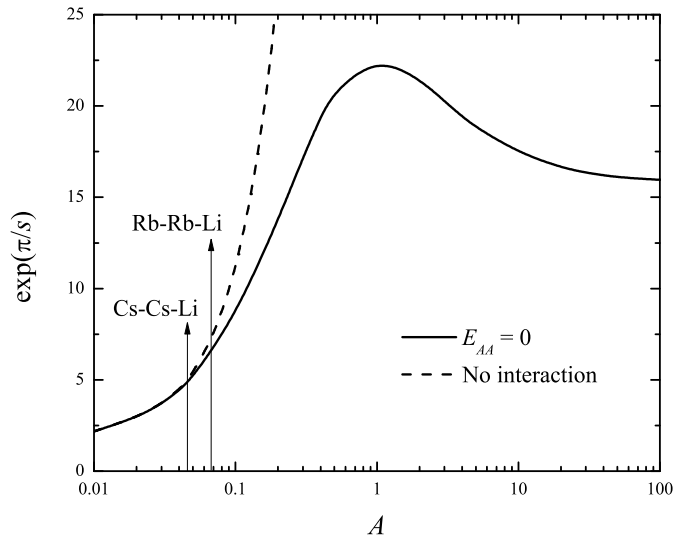,width=9cm} \caption[dummy0]{Scaling parameter $s$ as a
function of ${\cal A}=m_B/m_A$ for $E_{AA}=0$ and $E_{AB}=0$(resonant interactions), solid line,
and for the situation
where $E_{AB}=0$ but with no interaction between AA, dashed line. The arrows
show the corresponding mass ratios for $^{133}$Cs-$^{133}$Cs-$^{6}$Li and
$^{87}$Rb-$^{87}$Rb-$^{6}$Li.} \label{s}
\end{figure}

Inserting Eq.~\eqref{chi1a1} into Eq.~\eqref{chi2a1},  the set of coupled equations can be written as a single
transcendental equation:
\begin{eqnarray}
\frac{1}{\pi}\left(\frac{{\cal A}+1}{2{\cal A}}\right)^{3/2}  \sqrt{\frac
{2({\cal A}+1)}{{\cal A}+2}}\left({\cal A}I_1(s)
+\frac{2}{\pi}\sqrt{\frac {4{\cal A}}{{\cal A}+2}}I_2(s)I_3(s)\right)
=1 \ , \label{chi2a2}
\end{eqnarray}
where we have defined
\begin{eqnarray}
\label{I1}&I_1(s)=\int_{0}^{\infty}dz z^{-1+is}\,\,\textrm{log}\left[\frac{(z^2+1)({\cal A}+1)+2z}{(z^2+1)({\cal A}+1)-2z}\right]=\frac{2\pi}{s}\frac{\sinh(\theta_1 s-\frac{\pi}{2}s)}{\cosh(\frac{\pi}{2}s)},&\\
&I_2(s)=\int_{0}^{\infty}dz z^{-1+is}\,\,\textrm{log}\left[\frac{2{\cal A}(z^2+z)+{\cal A}+1}{2{\cal A}(z^2-z)+{\cal A}+1}\right]=
\frac{2\pi}{s}\frac{\sinh(\theta_2 s-\frac{\pi}{2}s)}{\cosh(\frac{\pi}{2}s)}\left(\frac{{\cal A}+1}{2{\cal A}}\right)^{is/2}&\\
&I_3(s)=\int_{0}^{\infty}dz z^{-1+is}\,\,\textrm{log}\left[\frac{2{\cal A}(1+z)+({\cal A}+1)z^2}{2{\cal A}(1-z)+({\cal A}+1)z^2}\right]=
\frac{2\pi}{s}\frac{\sinh(\theta_2 s-\frac{\pi}{2}s)}{\cosh(\frac{\pi}{2}s)}\left(\frac{{\cal A}+1}{2{\cal A}}\right)^{-is/2}.&
\end{eqnarray}
The angles are given by the equations $\tan^2\theta_1={\cal A}({\cal A}+2)$ and $\tan^2\theta_2=({\cal A}+2)/{\cal A}$ with
the conditions that $\pi/2<\theta_1,\theta_2<\pi$. For the special case of equal masses, i.e. ${\cal A}=1$, we have $\theta_1=\theta_2$, 
$I_1=I_2=I_3$ and
\begin{equation}
\left(\frac{1}{\pi}\sqrt{\frac{4}{3}}I_1(s)\right)+2\left(\frac{1}{\pi}\sqrt{\frac{4}{3}}I_1(s)\right)^2-1=0,
\end{equation}
for which the physically relevant solution is seen to be
\begin{equation}
\frac{1}{\pi}\sqrt{\frac{4}{3}}I_1(s)=\frac{1}{2}.
\label{sequal}
\end{equation}
Using Eq.~\eqref{I1}, we recover the celebrated Efimov equation for the scaling parameter, $s$, of
equal mass particles \cite{efi70,nie01,jen03,braaten2006}.
Another very interesting and relevant special case is when there is no interaction between the two $A$ particles, in
which case we can set $c_{AA}=0$ in the Eq.~\eqref{chi2a1}. The equation for the scale factor, Eq.~(\ref{chi2a2}), now simplifies and we
get
\begin{equation}\label{noAA}
\frac{{\cal A}}{\pi}\left(\frac{{\cal A}+1}{2{\cal A}}\right)^{3/2}
\sqrt{\frac{2({\cal A}+1)}{{\cal A}+2}}I_1(s)=1.
\end{equation}
This equation was first derived in Ref.~\cite{jen03} and later also discussed in Ref.~\cite{braaten2006}.
The derivation of Eq.~\eqref{sequal}) and \eqref{noAA} by using the asymptotic forms for the 
spectator functions reproduces the well known results for the scaling parameter $s$. In Fig.~\ref{s} 
we plot the scaling factors, $\exp(\pi/s)$, for the case when all three subsystems have resonant 
interaction which is the expression in Eq.~\eqref{chi2a2} valid for $E_{AA}=E_{AB}=0$ (solid line) 
and when there is no interaction in the $AA$ subsystem which is the expression in Eq.~\eqref{noAA} 
valid for $E_{AB}=0$ (dashed line). Our results are identical to the ones shown in  
Figs.~52 and 53 of Ref. \cite{braaten2006}.

What is important to notice is that for $m_A\gg m_B$ (${\cal A}\ll 1$), the
scaling factors are very similar, and both are much smaller than the
equal mass case where ${\cal A}=1$. We can therefore see that in the $AAB$ system with
heavy $A$ and light $B$, we should expect many universal three-body
bound states ($s$ large or equivalently $e^{\pi/s}$ small) 
{\it irrespective} of whether the heavy-heavy
subsystem is weakly or strongly interacting. Recent experiments
with mixtures of $^{6}$Li and $^{133}$Cs indicate that there could be
a resonance of the $^{6}$Li-$^{133}$Cs subsystem at a point where
the scattering length in the $^{133}$Cs-$^{133}$Cs system is close to 
zero, i.e. weak interaction in the $AA$ subsystem \cite{repp2013,tung2013}.

\begin{figure}[htb!]
\epsfig{file=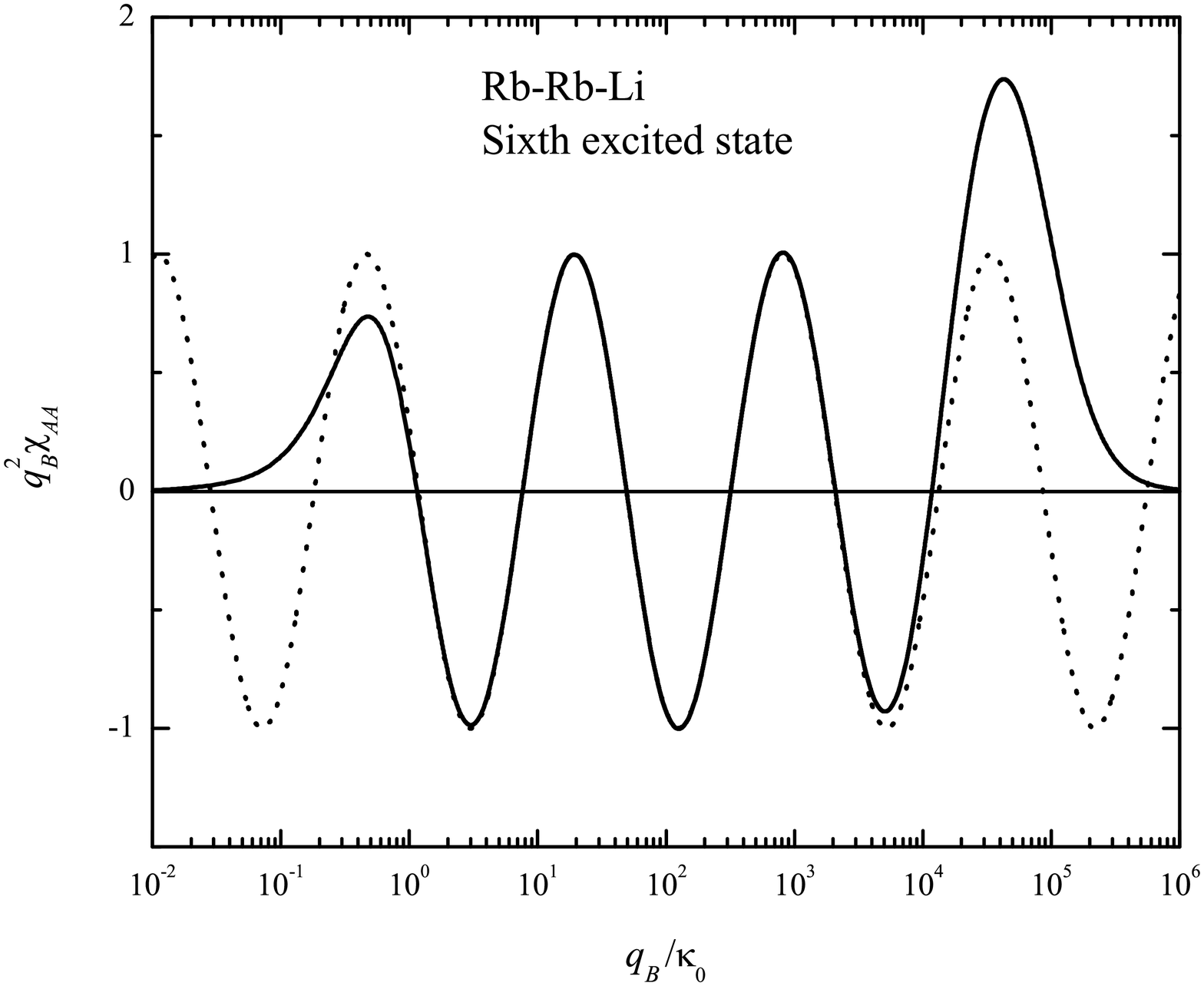,width=8cm}
\epsfig{file=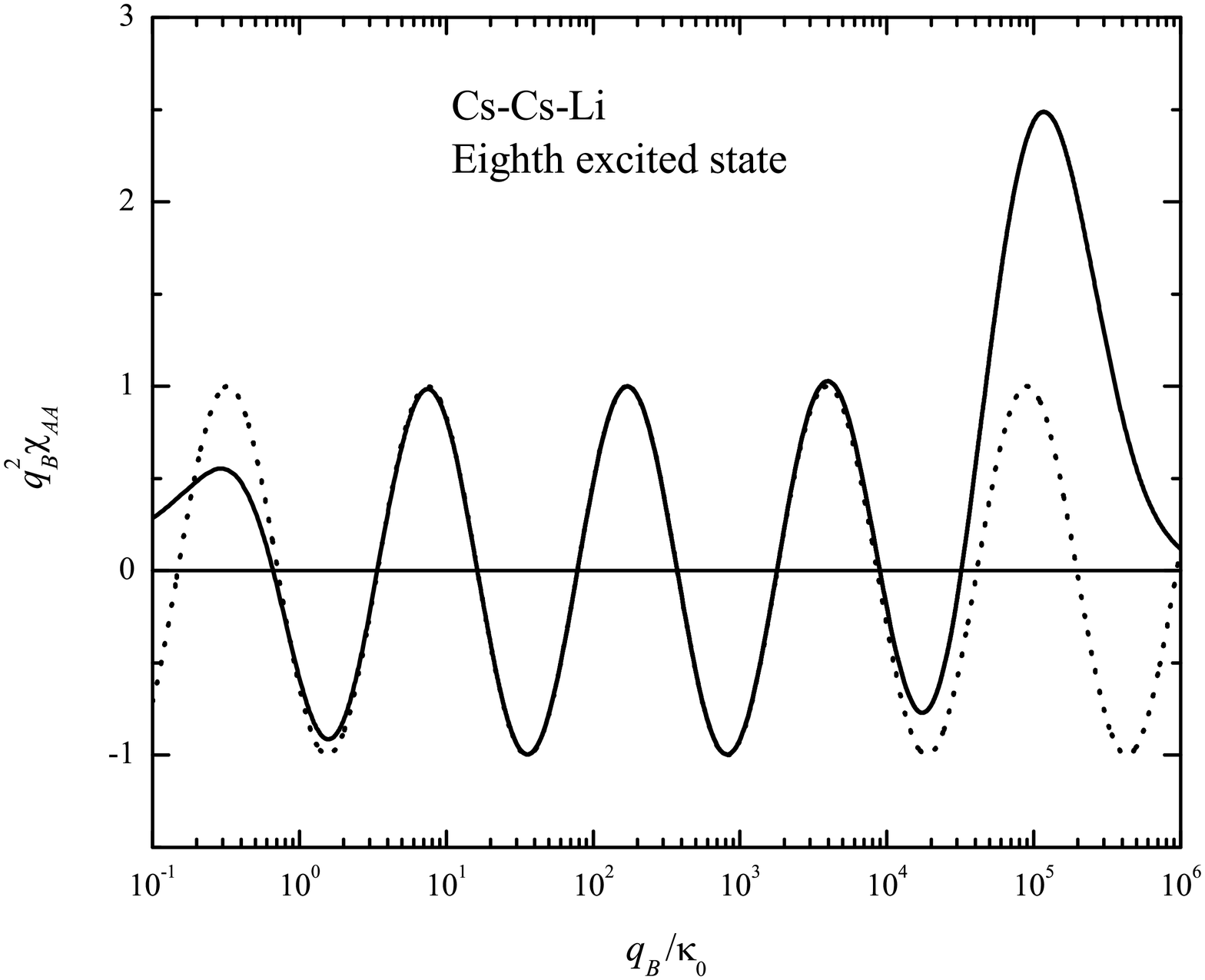,width=8cm}
\caption[dummy0]{{\bf Left:} $\chi_{AA}(q)$ of the sixth excited state
for $E_{AA}=E_{AB}=0$, $E_3=-8.6724\times10^{-12}E_0$, solution of the
coupled equations (\ref{chi1}) and (\ref{chi2}) (solid line),
compared with the asymptotic formula (\ref{sol}) for Rb-Rb-Li
molecule (dotted line). {\bf Right:} Same as left side for the eighth
excited state of Cs-Cs-Li, $E_3=-8.9265\times10^{-13}E_0$. Here we have 
defined $E_0=\hbar^2\mu^2/m_A$ and we work in units where $\hbar=m_A=\mu=1$ 
as explained in the text.} 
\label{chirb}
\label{figasym}
\end{figure}

Returning to Eqs.~\eqref{chi1a} and \eqref{chi2a}, there are two
solutions which are complex conjugates of each other, i.e. $z^{\pm\imath s}$.
Apart from an overall normalization, there is still a relative phase between 
these two independent solutions. We determine this phase by requiring 
that the wave function be zero at a certain momentum denoted $q^*$. 
This parameter is known as the three-body parameter 
\cite{nie01,braaten2006}. This is the 
momentum-space equivalent of the coordinate-space three-body parameter
which is now believed to be simply related to the van der Waals 
two-body interaction of the atoms in question 
\cite{berninger2011,naidon2011,chin2011,schmidt2012,wang2012,peder2012,naidon2012,wangwang2012,peder2013}.
In this case the
asymptotic form of the spectator functions becomes
\begin{equation}
\chi_{AA}(q)=c_{AA}\; q^{-2} \sin(s\;\log q/q^*) ~~~~{\text{and}}
~~~\chi_{AB}(q)=c_{AB}\; q^{-2} \sin(s\;\log q/q^*).
\label{chiasymp}
\end{equation}
Here we use $q$ to denote momentum and we see that our boundary condition
$\chi(q^*)=0$ is fulfilled.
The asymptotic form of the spectator function should be compared
with the solutions of the subtracted equations in the limit of large
momentum, constrained by the window $\kappa_0<<q_B<<\mu$, where 
$\kappa_0\equiv\sqrt{E_3}$. 
The spectator functions $\chi_{AA}(q)$ for Rb-Rb-Li and Cs-Cs-Li 
compared to the respective asymptotic formula are 
shown in figure \ref{figasym}. 
In the idealized limit where $\kappa_0=0$ and $\mu\to \infty$
the two curves would coincide. We can thus see the effect of finite 
value of these two quantities on each end of the plots.
The window of validity for the use of the asymptotic formulas, i.e., 
$\sqrt{E_3}<<q<<\mu$ can be clearly seen in these figures. 

\section{Asymptotic momentum density}\label{amomentum}
In this section we discuss the asymptotic momentum density for
$n(q_B)$, i.e. the single-particle momentum distribution for the 
$B$ particle. From Eqs.~\eqref{psiqb} and \eqref{nqaqb} we can split the 
momentum density into nine terms, which can be reduced to four considering 
the symmetry between the two identical particles $A$. This simplifies 
the computation of the momentum density to the form
\begin{equation}\label{4sum}
n(q_B)=\sum_{i=1}^4 n_i(q_B)
\end{equation}
where
\begin{eqnarray}
n_1(q_B)&=&\left|\chi_{AA}(q_B)\right|^2\int d^3p_B
 \frac{1}{\left(E_3+p_B^2+q_B^2\frac{{\cal A}+2}{4{\cal A}}\right)^2}\
 =\pi^2\frac{\left|\chi_{AA}(q_B)\right|^2 }{\sqrt{E_3+q_B^2\frac{{\cal A}+2}{4{\cal A}}}}
 , \\
n_2(q_B)&=& 2\int d^3p_B
 \frac{\left|\chi_{AB}(|\vec{p}_B-\frac{\vec{q}_B}{2}|)\right|^2}{\left(E_3+p_B^2+q_B^2\frac{{\cal A}+2}{4{\cal A}}\right)^2}
 =2\int d^3q_A
 \frac{\left|\chi_{AB}(q_A)\right|^2}{\left(E_3+q_A^2+\vec q_A\cdot \vec
q_B+q_B^2\;\frac{{\cal A}+1}{2{\cal A}}\right)^2}
 \\
n_3(q_B)&=& 2\;\chi^{\ast}_{AA}(q_B)\int d^3p_B
 \frac{\chi_{AB}(|\vec{p}_B-\frac{\vec{q}_B}{2}|)}{\left(E_3+p_B^2+q_B^2\frac{{\cal A}+2}{4{\cal A}}\right)^2}+c.c.\label{n3}\\
n_4(q_B)&=& \int d^3p_B
 \frac{\chi^{\ast}_{AB}(|\vec{p}_B-\frac{\vec{q}_B}{2}|)\chi_{AB}(|\vec{p}_B+\frac{\vec{q}_B}{2}|)}
 {\left(E_3+p_B^2+q_B^2\frac{{\cal A}+2}{4{\cal A}}\right)^2}+c.c. \label{n4}
\end{eqnarray}

\begin{figure}[htb!]
\epsfig{file=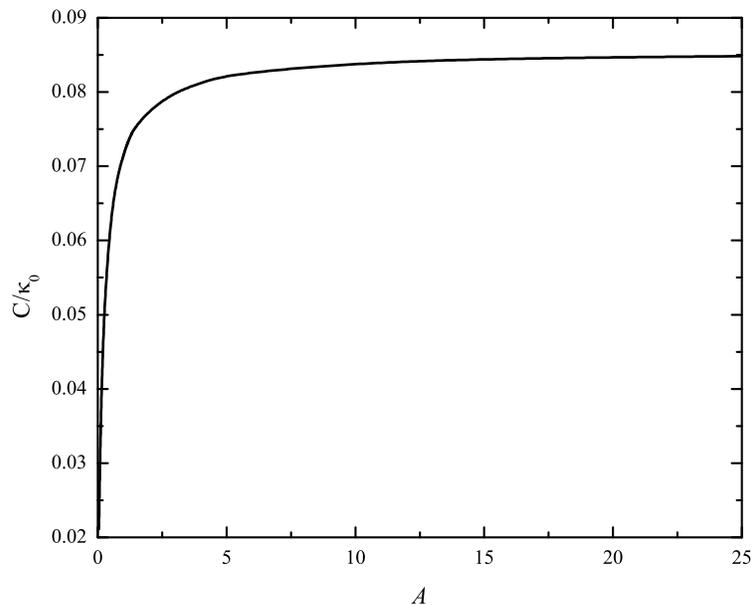,width=10cm}
\caption{$C/\kappa_0$ for mass ratios ranging from ${\cal A}=6/133$ to ${\cal A}=25$. These 
results should be multiplied by the factor $3(2\pi)^3$ in order to be compared with 
Ref.~\cite{castin2011}.}
\label{figc}
\end{figure}

The leading order term $\frac{C}{q_B^4}$ comes only from $n_2$ and the constant 
$C$ is simply given by $C=\frac{8{\cal A}^2}{({\cal A}+1)^2}\int d^3q_A
\left|\chi_{AB}(q_A)\right|^2$. This formula gives $C/\kappa_0=0.0274$ for 
$^{133}$Cs-$^{133}$Cs-$^6$Li and $C/\kappa_0=0.0211$ for $^{87}$Rb-$^{87}$Rb-$^6$Li. 
For ${\cal A}=1$ we obtain $3(2\pi)^3C/\kappa_0=53.197$, to be compared with the ``exact'' 
value, 53.097, obtained in Ref.~\cite{castin2011}. The factor $3(2\pi)^3$comes 
from the difference in choice of normalization. 
In Fig.~\ref{figc} we plot the value 
of $C/\kappa_0$ for mass ratios ranging from ${\cal A}=6/133$ to ${\cal A}=25$. 
The increase is very rapid until ${\cal A}\sim 5$ beyond which an almost constant
value is reached.
Note that we have used the second excited state to perform these calculations for $C$. 
This can explain the small discrepancy between the numerical and ``exact'' result, 
which was calculated for an arbitrary high excited state.

\subsection{Analysis of subleading terms}
In Ref.~\cite{castin2011} it was shown that the non-oscillatory term of 
order $q_B^{-5}$ coming from $n_1$ to $n_4$ cancels for ${\cal A}=1$ (equal masses). 
Here we will demonstrate that 
this conclusion does not hold for general ${\cal A}\neq1$. Below we will consider
the $E_3\to 0$ limit, i.e. the three-body energy is assumed to be negligible, since 
we are interested in the imprint of Efimov states on the momentum distribution for 
excited Efimov states that are very extended and do not feel any short-range effects
(as encoded in the three-body parameter, $q^*$, discussed above).

$\bullet$ $n_1$: Upon inserting \eqref{chiasymp} in the momentum distributions we get
for $n_1(q_B)$, the following asymptotic expression
\begin{eqnarray}
n_1(q_B)&\to&
 2\pi^2\sqrt{ \frac{{\cal A}}{{\cal A}+2}}\;\frac{\left|\chi_{AA}(q_B)\right|^2 }{q_B }
\to 2\pi^2\; \left|c_{AA}\right|^2\;  \sqrt{ \frac{{\cal A}}{{\cal
A}+2}}\frac{\left|\sin(s\;\ln q_B/q^*)\right|^2}{q_B^5} .
\end{eqnarray}
Averaging out the oscillating part yields $1/2$ and we have 
$\langle n_1(q_B)\rangle=\frac{\pi^2}{q_B^5}\; \left|c_{AA}\right|^2\;  \sqrt{ \frac{{\cal A}}{{\cal A}+2}}$.

$\bullet$ $n_2$: For large $q_B$ this term
becomes
\begin{align}
n_2(q_B)= &2\int d^3q_A
 \frac{\left|\chi_{AB}(q_A)\right|^2}{\left(q_A^2+\vec q_A\cdot \vec
q_B+q_B^2\;\frac{{\cal A}+1}{2{\cal A}}\right)^2} =
\frac{8{\cal A}^2}{q_B^4\;({\cal A}+1)^2}\int d^3q_A\;\left|\chi_{AB}(q_A)\right|^2 &\nonumber\\
&+ \int d^3q_A \left|\chi_{AB}(q_A)\right|^2\left[
 \frac{2}{\left(q_A^2+\vec q_A\cdot \vec
q_B+q_B^2\;\frac{{\cal A}+1}{2{\cal A}}\right)^2}-\frac{8{\cal
A}^2}{({\cal A}+1)^2}\frac{1}{q_B^4}\right],& \label{n2asym}
\end{align}
where we retain a sub-leading part since it is of the same order as
the leading order of the other terms. This integral can be solved 
analytically (see App.~\ref{appendix1}) and the final expression is
\begin{equation}
\left<n_2(q_B)\right>=-\frac{8\pi^2 \left|c_{AB}\right|^2}{q_{B}^{5}} \frac{{\cal
A}^3({\cal A}+3)}{({\cal A}+1)^3 \sqrt{{\cal A}({\cal A}+2)}}.
\end{equation}
The special case ${\cal A}=1$ yields $\left<n_2(q_B)\right>=-4\pi^2 \left|c_{AB}\right|^2/(\sqrt{3}q_{B}^{5})$.

$\bullet$ $n_3$:
The term for $n_3$ is considerably more complicated since it involves an angular integral. The details can be 
found in App.~\ref{appendix1} and the final result is
\begin{equation}
\langle n_3(q_B)\rangle= \frac{4 \pi^2 c_{AA}\;c_{AB}}{q_B^5 \cosh\left(\frac{s\pi}{2}\right)}\left\{\sqrt{\frac{{\cal A}}{{\cal A}+2}}\cos\left(s\ln \sqrt{\frac{{\cal A}+1}{2{\cal A}}}\right)\cosh\left[s\left(\frac{\pi}{2}-\theta_3\right)\right]+\sin\left(s\ln \sqrt{\frac{{\cal A}+1}{2{\cal A}}}\right)\sinh\left[s\left(\frac{\pi}{2}-\theta_3\right)\right]\right\} \ ,
\label{eq:}
\end{equation}
where $\tan\theta_3=\sqrt{\frac{{\cal A}+2}{{\cal A}}}$ for $0\leq\theta_3\leq\pi/2$. The special case ${\cal A}=1$ yields $\theta_3=\pi/3$ and $\langle n_3(q_B)\rangle= 4 \pi^2 |c_{AA}|^2 \cosh\left(\frac{s\pi}{6}\right) /\left(q_B^5 \sqrt{3} \cosh\left(\frac{s\pi}{2}\right)\right)$.

$\bullet$ $n_4$: The $n_4$ terms is also complicated by angular integrals and again we refer to App.~\ref{appendix1} for details.
The result is
\begin{equation}
\left\langle n_4(q_B) \right\rangle=\frac{8\pi^2 |c_{AB}|^2 {\cal A}^2 }{s\; q_B^5\; \cosh\left(\frac{s\pi}{2}\right)}\left\{ \sinh\left[s\left(\frac{\pi}{2}- \theta_4\right) \right]-\frac{s\; {\cal A} }{\sqrt{{\cal A}({\cal A}+2)} ({\cal A}+1)}\cosh\left[s\left(\frac{\pi}{2}- \theta_4\right) \right]\right\} \ ,
\end{equation} 
where $\tan\theta_4=\sqrt{{\cal A}({\cal A}+2)}$ for $0\leq\theta_4\leq\pi/2$. The special case ${\cal A}=1$ yields $\theta_4=\pi/3$ and $\langle n_4(q_B)\rangle= 8 \pi^2 |c_{AA}|^2\left[\sinh\left(\frac{s\pi}{6}\right)-s/(2\sqrt{3})\cosh\left(\frac{s\pi}{6}\right) \right]  /\left[s\; q_B^5\;  \cosh\left(\frac{s\pi}{2}\right)\right]$.

\begin{figure}[htb!]
\epsfig{file=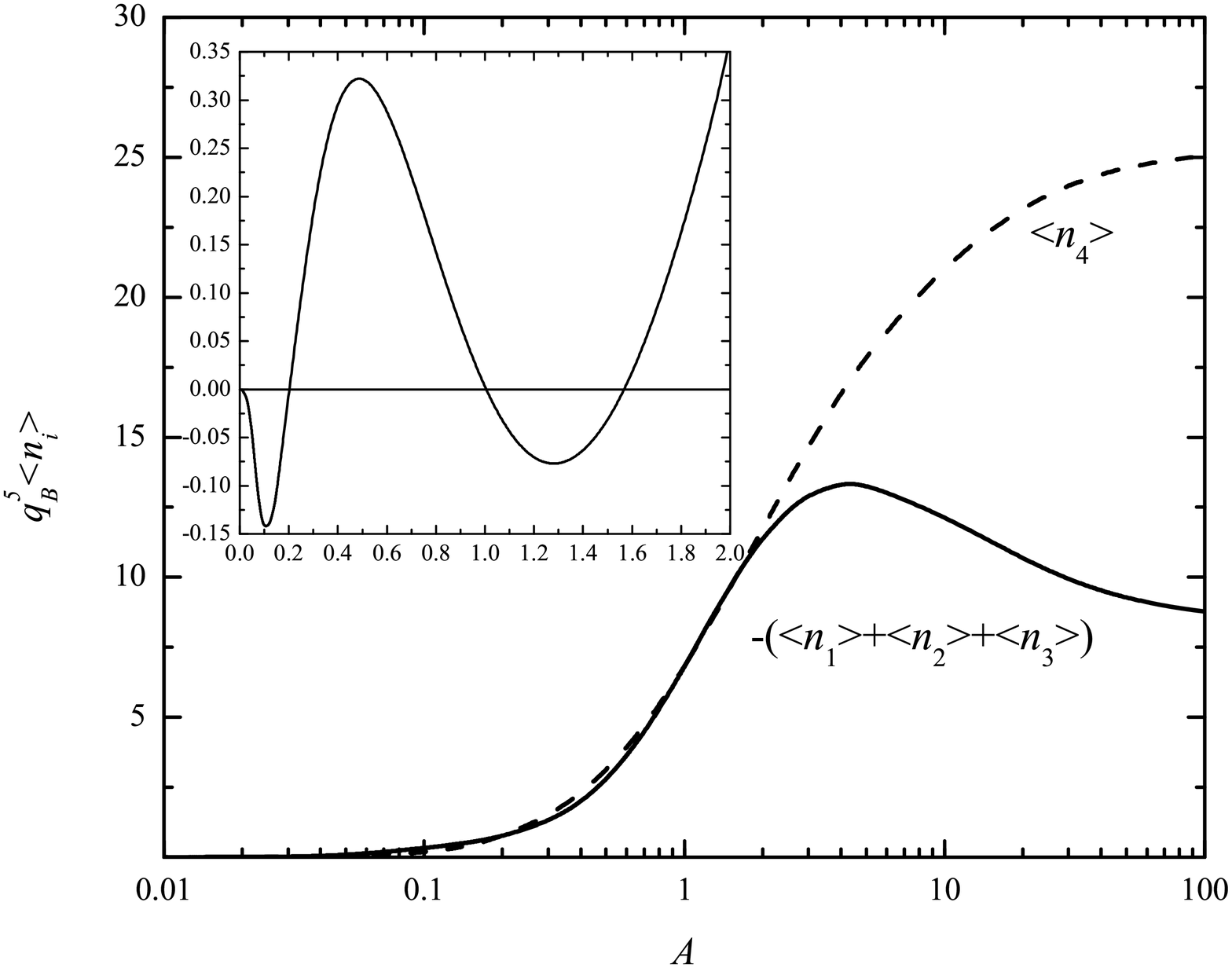,width=8cm}
\epsfig{file=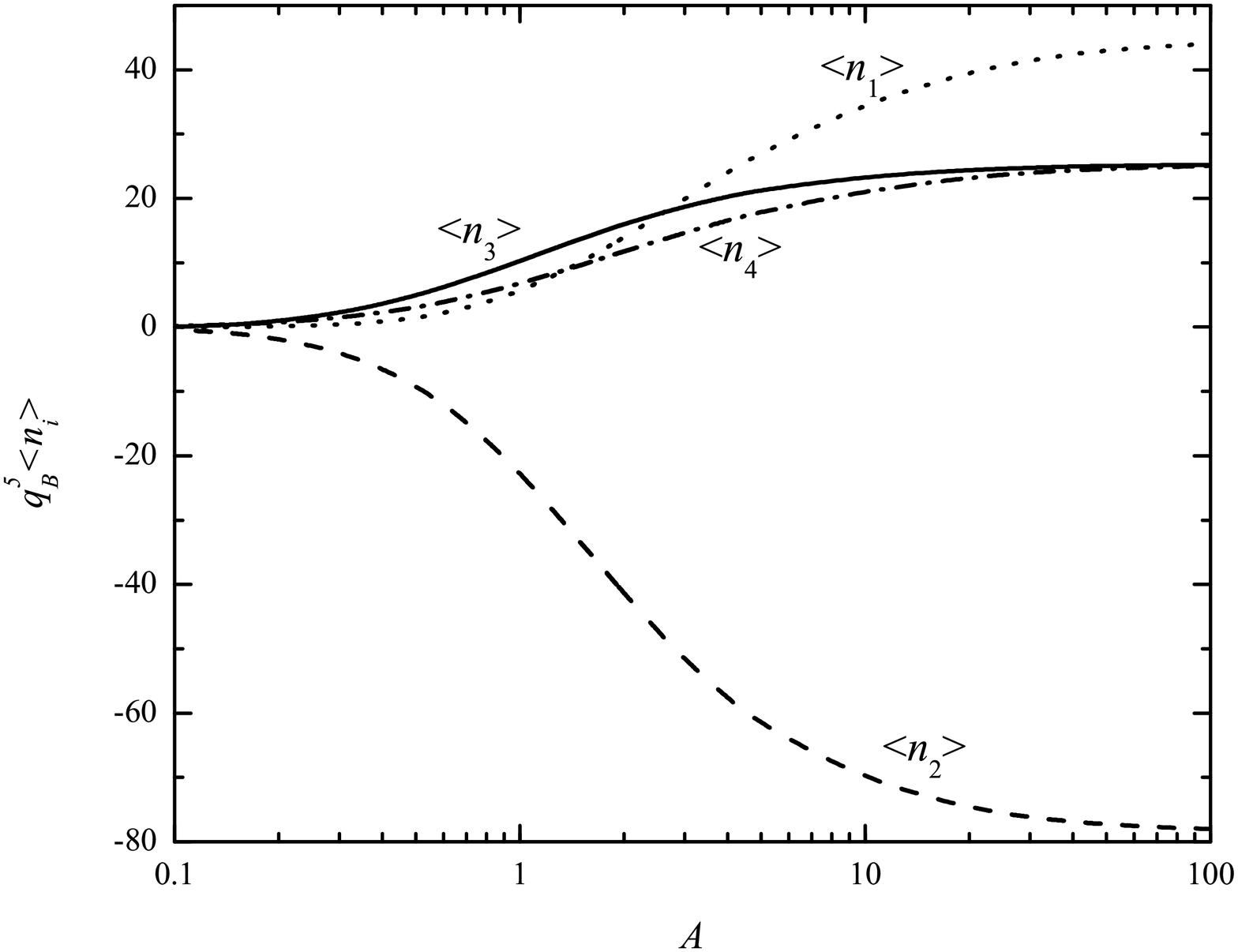,width=8cm}
\caption{(Left) Non-oscillatory contributions for $n_1+n_2+n_3$, and $n_4$ as a function of the mass 
ratio ${\cal A}$. Their sum, showed in the inset on the left, cancels exactly for ${\cal A}=$ 0.2, 1 and 1.57. 
(Right) The individual contributions $n_1$, $n_2$, $n_3$, and $n_4$ as function of ${\cal A}$.}
\label{cancel_nonosc}
\end{figure}

It is important to note here that the asymptotic forms for the spectator functions have been used 
in the integrals where the integration is being performed from 0 to $\infty$. This may a priori 
cause problems for small momenta. However, a numerical 
check shows that the different behaviour of the spectator functions at low momenta contributes 
only in an order higher than $q_B^{-5}$ to the integrals. This procedure is the same as 
that used in Ref.~\cite{castin2011}.

We now have analytic expressions for all the four terms in Eq.~\eqref{4sum}.  The ratio between the coefficients $c_{AA}$ 
and $c_{AB}$ is given by 
Eq.~\eqref{chi1a1} which can be used to eliminate one of these normalization factors. The other one can be determined
from the overall normalization of the wave function which we will not be concerned with here and we will merely set 
$c_{AB}=1$ from now on. 
In Fig.~\ref{cancel_nonosc} we plot the contribution $-(n_1+n_2+n_3)$ and $n_4$ as a 
function of mass ratio ${\cal A}$ on the left-hand side and the 
individual contributions of $n_1$, $n_2$, $n_3$, and $n_4$ separately on the right-hand side. 
What is immediately seen is that for ${\cal A}=1$ we reproduce the 
result of Ref.~\cite{castin2011}, i.e. that the $q_{B}^{-5}$ non-oscillatory term cancels. However, 
for general ${\cal A}$ this is not the case and one should expect also a $q_{B}^{-5}$ term in the 
asymptotic momentum distribution for systems with two identical and a third particle when three-body 
bound states are present. This is the main result of our paper and it demonstrates that non-equal 
masses will generally influence not only the value of the contact parameter attributed to three-body
bound states but also the functional form of the asymptotic momentum tail. Curiously, there is an 
oscillatory behavior around ${\cal A}\sim 1$ of the sum of all contributions. This is shown in the
inset of Fig.~\ref{cancel_nonosc} where we see zero-crossings at ${\cal A}=0.2$, $1$, and $1.57$.
It seems quite clear that the oscillatory terms that all depend on the scale factor, $s$, are to
blame for this interesting behavior, but we have not found an easy analytic explanation for it. 
What makes this interesting is the fact that if we take ratios of typical isotopes of alkali atoms 
like Li, Na, K, Rb, and Cs, one can get rather close to 0.2 or 1.57. For instance, taking one $^{133}$Cs
and two $^{85}$Rb yields ${\cal A}=1.565$, while one $^{7}$Li atom and two $^{39}$K atoms yields
${\cal A}=0.179$. These interesting ratios are thus close to experimentally accessible species.

\begin{figure}[htb!]
\epsfig{file=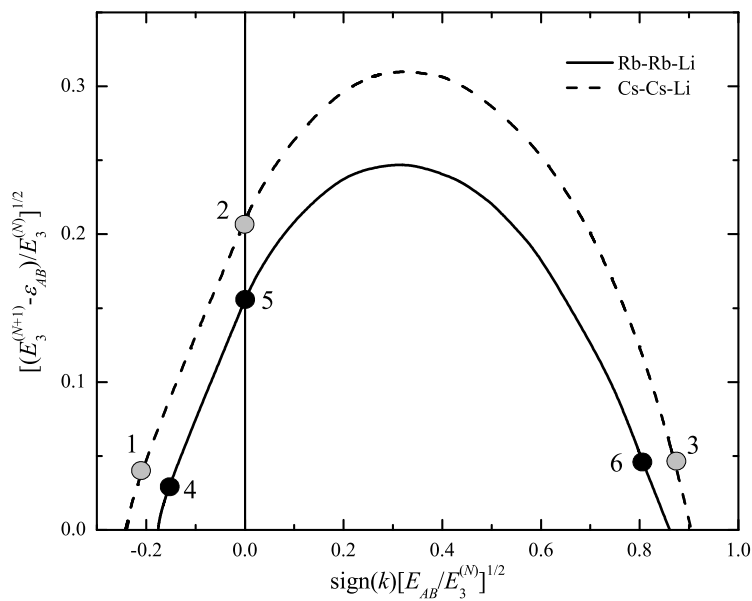,width=10cm}
\caption{
Square-root of the ratio of the $(N+1)$th three-body state (measured from threshold $\epsilon_{AB}$) to the
$N$th state plotted as function of the square-root of the ratio of 
$E_{AB}$ and the energy of the $N$th three-body bound state. Note the factor $\textrm{sign}(k)$ which 
indicates whether the two-body system has a bound ($k=+1$) or virtual ($k=-1$) state. This sign 
is consistent with the convention introduced in Eqs.~\eqref{tau1} and \eqref{tau2}.
The limit cycle, which should be in principle reached 
for $N\rightarrow\infty$, is achieved very fast so that the curve has been constructed 
using $N=2$. 
$E_{AB}$ is the Cs-Li or Rb-Li two-body energy (Cs-Cs and Rb-Rb 
two-body energies are zero). The negative and positive parts refer, respectively, to virtual and 
bound $AB$ states, such that $\epsilon_{AB}=0$ and $\epsilon_{AB}\equiv E_{AB}$, respectively, 
on the negative and positive sides. The circles labelled from 1 to 6 mark the points where the 
momentum distributions have been calculated.}
\label{bell}
\end{figure}

\section{Numerical examples}\label{results}
We now provide some numerical examples of momentum distributions for the experimentally 
interesting systems with large mass ratios. We will focus on
$^{133}$Cs-$^{133}$Cs-$^6$Li and $^{87}$Rb-$^{87}$Rb-$^6$Li. 
Here we will investigate two extreme possibilities: (i) the heavy-heavy subsystems, 
i.e. $^{133}$Cs-$^{133}$Cs and $^{87}$Rb-$^{87}$Rb, 
have a two-body bound state at zero energy
and (ii) the opposite limit where they do not interact. 
In the first case the heavy atoms are 
at a Feshbach resonance with infinite scattering length, 
while in the second case they are far from resonance and we assume a 
negligible background scattering length.
As was recently 
demonstrated for the $^{133}$Cs-$^{6}$Li mixture, there are Feshbach resonances 
in the Li-Cs subsystem at positions where the Cs-Cs scattering length is non-resonant
\cite{repp2013,tung2013}. 
While this does not automatically imply that the Cs-Cs channel
can be neglected, we will make the assumption (ii) here. 
The formalism can be modified in a 
straightforward manner to also include interaction in the heavy-heavy subsystem.

As before we denote the system $AAB$, where $A$ refers to the identical (bosonic) atoms,
$^{133}$Cs or $^{87}$Rb, and $B$ to $^6$Li. 
By solving Eq.~\eqref{chi2a2}, one finds $s(6/133)=2.00588$ and $s(6/87)=1.68334$ when
assuming that all three subsystems have large scattering lengths (solid line in Fig.~\ref{s}).
The situation where the interaction between the two 
identical particles is turned off is shown by the dashed line in Fig.~\ref{s}. 
In this case, $s({\cal A})$ was calculated from Eq.~\eqref{chi2a1} by setting $c_{AA}=0$. This 
yields $s(6/133)=1.98572$ and $s(6/87)=1.63454$.

\begin{figure}[htb!]
\epsfig{file=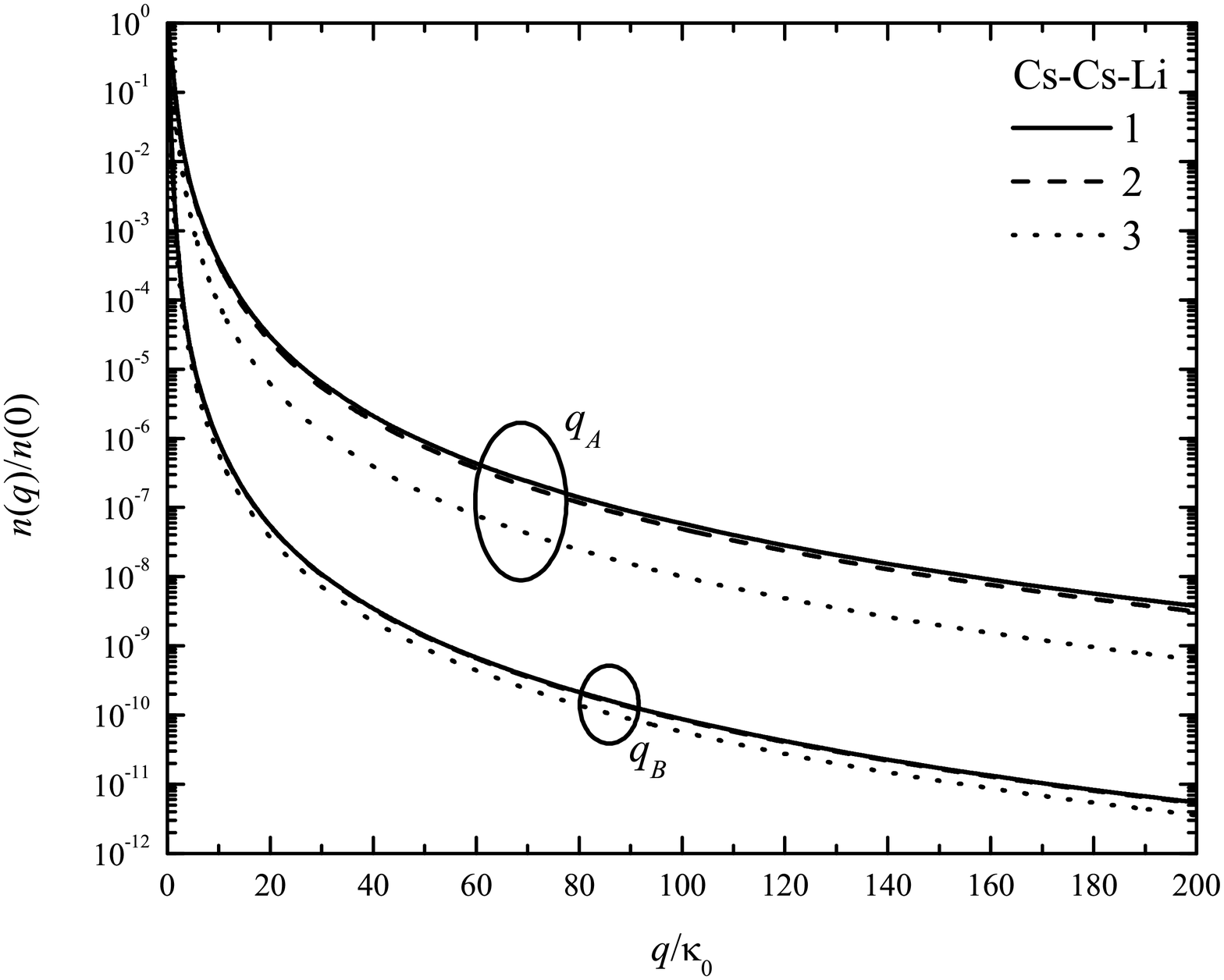,width=8cm}
\epsfig{file=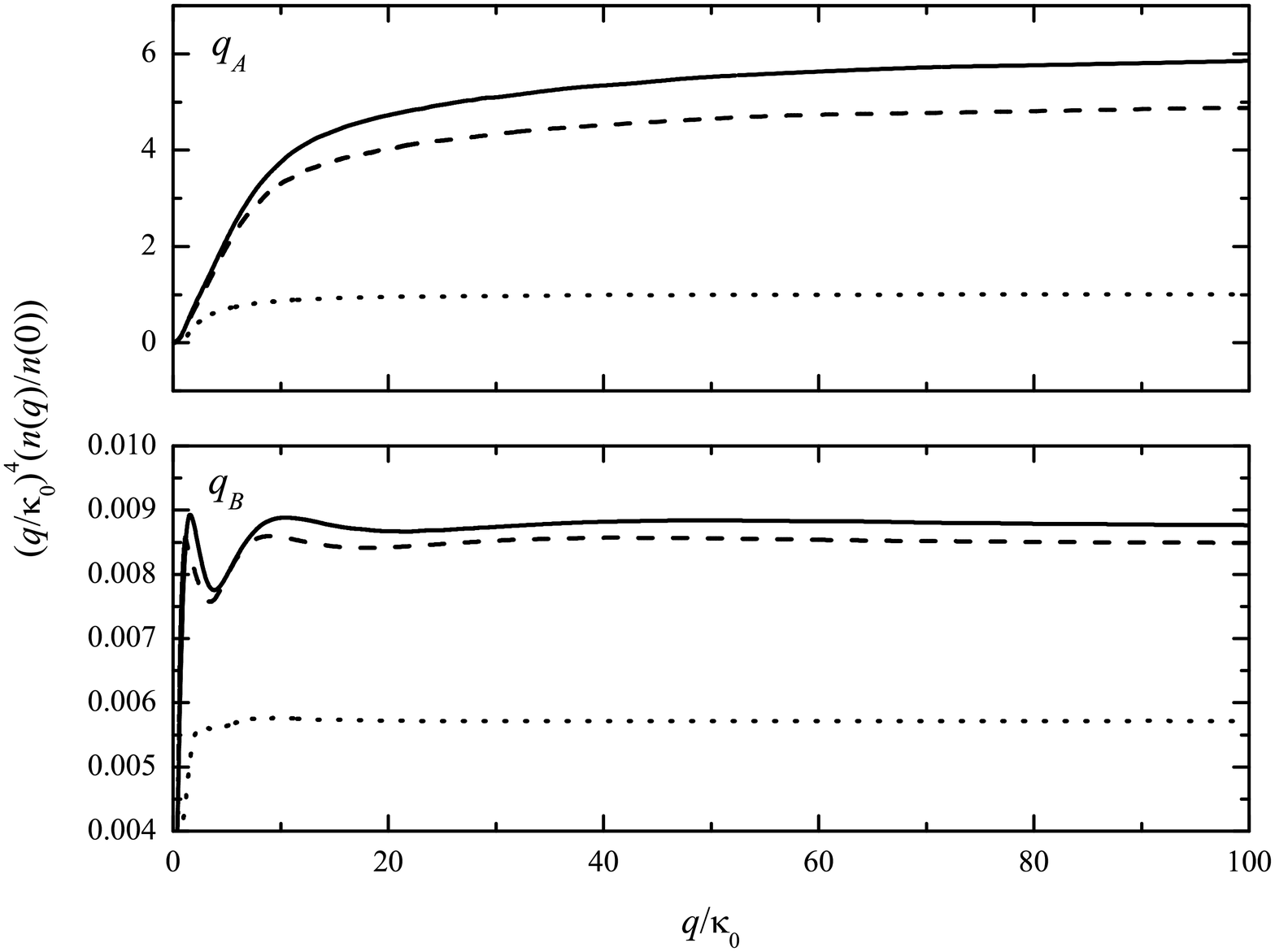,width=8.5cm}
\caption{
Left: Momentum distribution for the second excited state as a function of the relative
momentum of one $^{133}$Cs, $q_A$, or $^6$Li, $q_B$, to the center-of-mass of the remaining pair 
$^{133}$Cs-$^6$Li or $^{133}$Cs-$^{133}$Cs. The solid, dashed and dotted lines were calculated 
for the two- and three-body energies satisfying the ratios indicated by the points 1 to 3 in figure 
\ref{bell}. The circles show the set of curves related to $q_A$ or $q_B$. Right: Same curves as on the 
left side multiplied by $q^4$, which shows explicitly the leading decay $1/q^4$.
}
\label{123Cs}
\end{figure}

We first consider the binding energies. Assuming that the Cs-Cs and Rb-Rb two-body energies 
are zero, we have, for a system satisfying the universality condition $a\gg r_0$, that any 
observable should be a function of the remaining two- and three-body scales, which can be 
conveniently chosen as $E_3^{(N)}$ and $E_{AB}$ (the Cs-Li or Rb-Li two-body energy). Here 
$N$ denotes the $N$th consecutive three-body bound state with $N=0$ being the lowest one. 
Thus, the energy of an $N+1$ state can be plotted in terms of a scaling function relating 
only $E_{AB}$ and the previous state. The limit cycle, which should be in principle reached 
for $N\rightarrow\infty$, is achieved rapidly so that we can construct 
the curve shown in Fig.~\ref{bell} using $N=2$ \cite{wilson,yamashita2002,frederico2012}. The negative and positive parts of the 
horizontal axis refer, respectively, to virtual and bound two-body $AB$ states. The circles
labelled from 1 to 6 mark the points where the momentum distributions have been calculated. 
The points 1 and 4 represent the Borromean case, the points 2 and 5 are the ``Efimov situation'' 
and in points 3 and 6 $AB$ is bound. 

\begin{figure}[htb!]
\epsfig{file=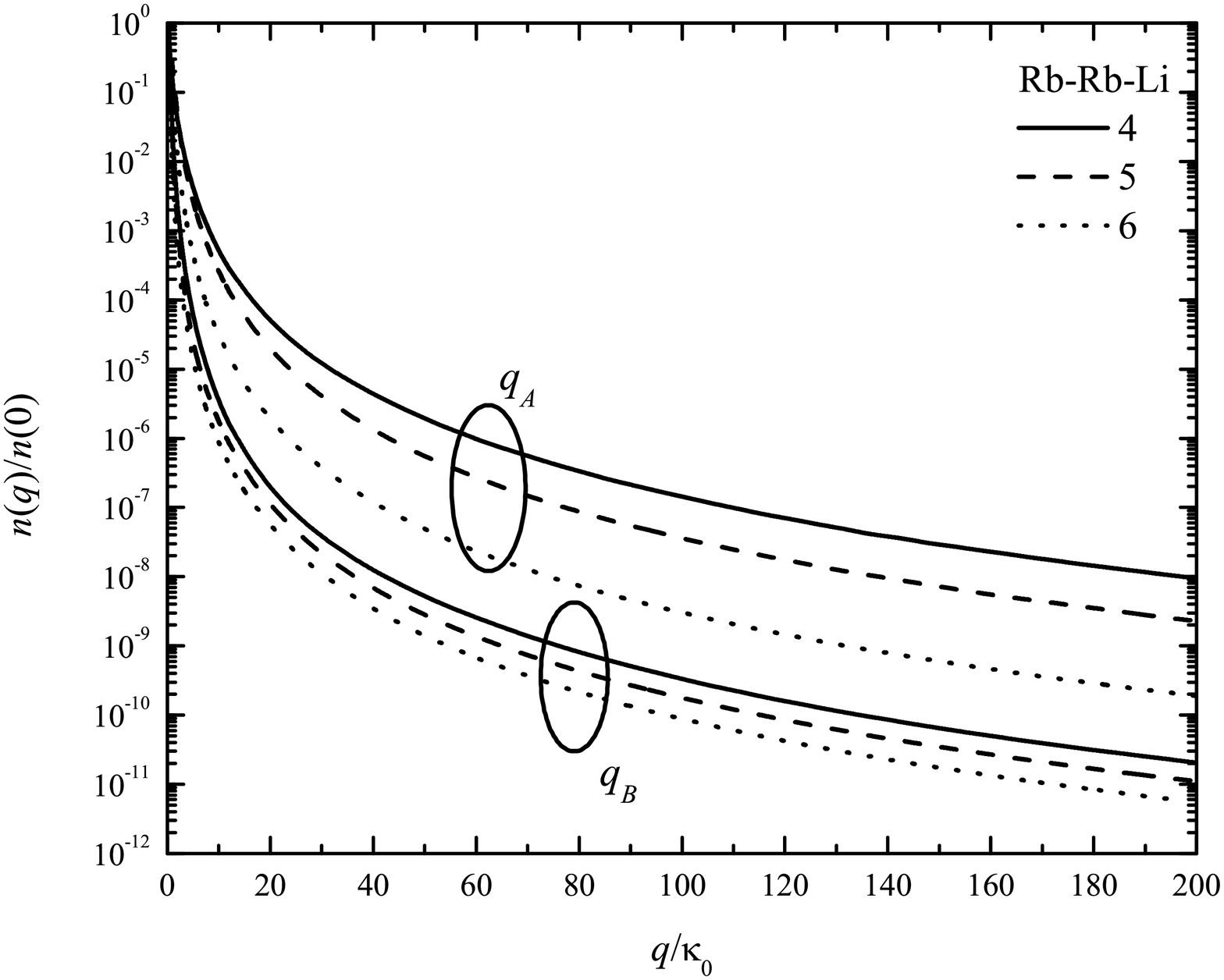,width=8cm}
\epsfig{file=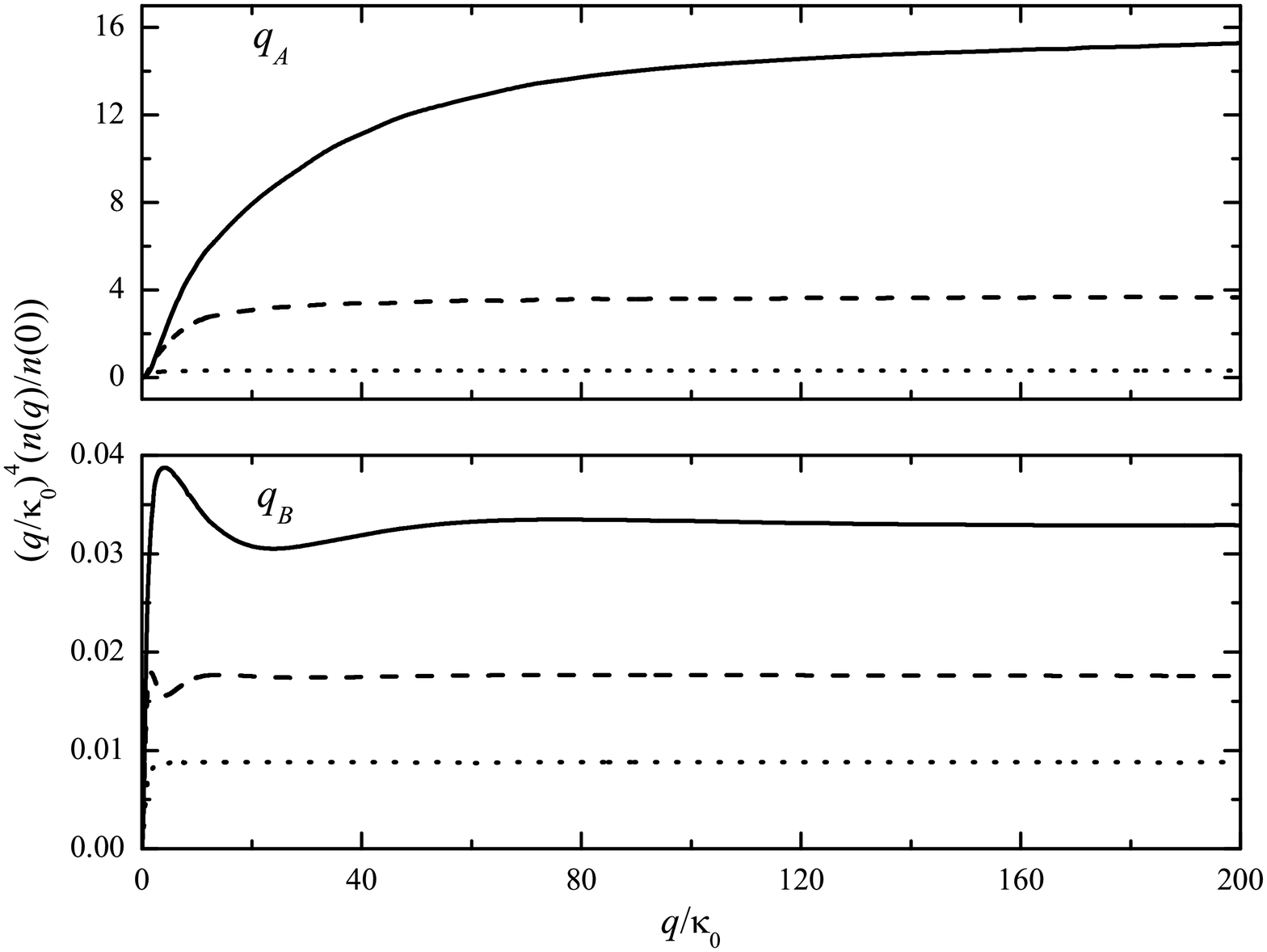,width=8.5cm}
\caption{
Momentum distribution for the second excited state as a function of the relative
momentum of one $^{87}$Rb, $q_A$, or $^6$Li, $q_B$, to the center-of-mass of the remaining pair 
$^{87}$Rb-$^6$Li or $^{87}$Rb-$^{87}$Rb. The solid, dashed and dotted lines were calculated 
for the two- and three-body energies satisfying the ratios indicated by the points 4 to 6 in figure 
\ref{bell}. The circles show the set of curves related to $q_A$ or $q_B$. Right: same curves of the 
left side multiplied by $q^4$, which shows explicitly the leading decay $1/q^4$.
}
\label{123Rb}
\end{figure}

Figures~\ref{123Cs} and \ref{123Rb} give the momentum distributions of the second excited states
for the energy ratio $\sqrt{E_{AB}/E_3}$ given by the points labeled from 1 to 6 in 
Fig.~\ref{bell}. According to our previous calculations \cite{nuclearphys04}, for fixed 
three-body energy the size of the system increases as the number of bound 
two-body subsystems increase. Thus, it seems reasonable that the momentum distribution for the Borromean case (point 1)
decreases slower. This behavior is clearly seen on the left side of 
figures \ref{123Cs} and \ref{123Rb}. The distance of one atom to the center-of-mass of 
the other two is much larger for $^6$Li than for $^{133}$Cs or $^{87}$Rb, due to the large 
difference of the masses, such that the decrease of the momentum distribution for the heavier 
atom, $q_A$ set, decreases much slower than that for the lighter one, $q_B$ set. This also reflects 
on the momentum from which the leading-order decay $1/q^4$ start to be dominant. This difference 
becomes evident on the right side of figures \ref{123Cs} and \ref{123Rb}, where we plotted $q^4n(q)$.
Thus the $q^4$ term is dominant above $(20-40)\kappa_0$ for $q_b$ and much slower for $q_A$ at about
$(60-100)\kappa_0$.

\begin{figure}[htb!]
\epsfig{file=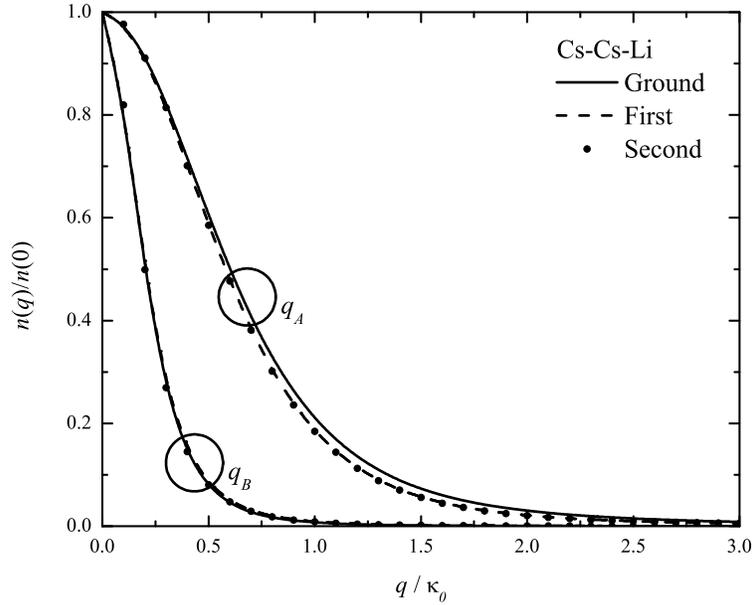,width=10cm}
\caption[dummy0]{Rescaled momentum distribution for the ground, first and second excited
states as a function of the relative momenta of $^{133}$Cs to the center-of-mass of the pair
$^{6}$Li-$^{133}$Cs, $q_A$, and $^6$Li to the center-of-mass of the pair
$^{133}$Cs-$^{133}$Cs, $q_B$. The subsystem binding energies are all set to zero.
Normalization to unity at zero momentum.}
\label{scalingCs}
\end{figure}

\begin{figure}[htb!]
\epsfig{file=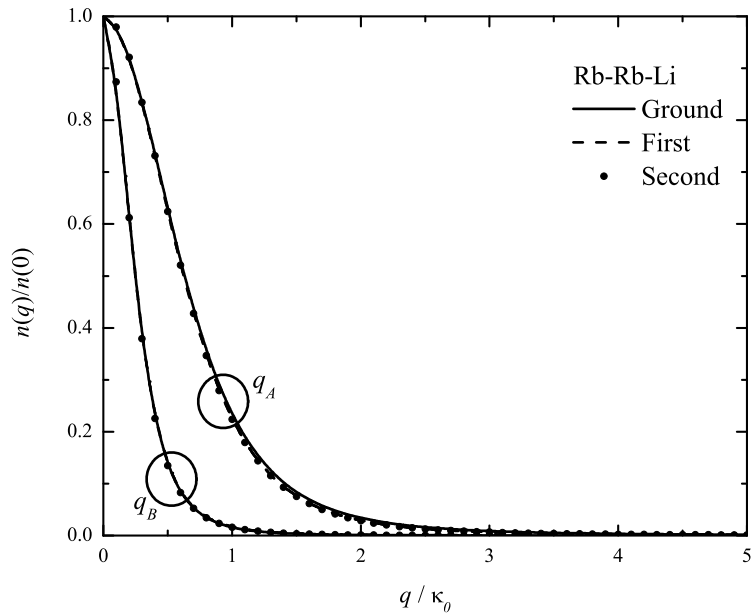,width=10cm}
\caption[dummy0]{Rescaled momentum distribution for the ground, first, second and third excited
states as a function of the relative momenta of $^{87}$Rb to the center-of-mass of the pair
$^{87}$Rb-$^{6}$Li, $q_A$, and $^6$Li to the center-of-mass of the pair
$^{87}$Rb-$^{87}$Rb, $q_B$. The subsystem binding energies are all set to zero.
Normalization to unity at zero momentum.}
\label{scalingRb}
\end{figure}

Figures~\ref{scalingCs} and \ref{scalingRb} show the rescaled momentum distributions for the
ground, first and second excited states. In these figures, the subsystem energies 
were chosen to zero, corresponding to the transition point to a Borromean configuration. 
In this situation, the only low-energy scale is $E_3$ (remember that the
high-momentum scale is $\mu=1$). Therefore, in units in which
$\mu=1$, to achieve a universal regime, in principle, to wash-out
the effect of the subtraction scale, $\mu$, we have to go to a highly excited state (see, 
for instance, Fig. (\ref{figasym}) and the comments inside the text associated to it).  
However, a universal low-energy regime of $n(q_B)/n(q_B=0)$ is seen for momentum of 
the order of $\sqrt{E_3}$, even for the ground state which is smaller 
than excited states. Thus, in practice, the universal 
behavior of the momentum distribution is approached rapidly.

\section{Conclusions and Outlook}\label{diss}
In the current work we have calculated the single-particle momentum distribution of 
systems consisting of two identical bosonic particles and a third particle of a different
kind with short-range interaction in the regime where three-body bound states and the 
Efimov effect occurs. We analytically
calculate the asymptotic momentum distribution as a function of the mass ratio and 
find that the functional form is sensitive to this ratio. In the case of equal mass
we reproduce the results of Ref.~\cite{castin2011}, i.e. that the leading term has 
a $q^{-4}$ tail while the subleading contribution is $q^{-5}$ times a log-periodic 
oscillatory function that is characteristic of the Efimov effect and that depends 
on the scale factor (and thus on the mass ratio) of the Efimov states. In particular,
we find that for general mass ratios, there is a non-oscillatory $q^{-5}$ contribution
which appears to only vanish (and leave the oscillatory contribution behind) when 
the mass ratio is 0.2, 1, or 1.57.

To examplify our study, we consider $^{133}$Cs-$^{133}$Cs-$^{6}$Li and 
$^{87}$Rb-$^{87}$Rb-$^{6}$Li where we numerically determine the coefficient of the 
$q^{-4}$ tail which is the two-body contact parameter introduced by Tan \cite{tan2008}.
For these examples, we also numerically determine the momentum distributions of excited
Efimov trimers for both the heavy and light components. Our numerical results demonstrate
that the momentum distributions of ground, first, and second excited Efimov trimers 
approach universal behavior very fast at large but also at small momentum, indicating that 
one does not need to go to highly excited (and numerically challenging) three-body states
in order to study the universal behavior of Efimov states in momentum space. 
Recent experiments have succesfully measured the momentum distribution of ultracold
atomic gases using time-of-flight and mapping to momentum space \cite{stewart2010}
and Bragg spectroscopy \cite{stewart2010,kuhnle2010,wild2012}. Observing a
constant $1/k^5$ contribution in a system with non-equal mass three-body states is
a considerable challenge since one needs to first subtract the leading-order $1/k^4$
contribution. The $1/k^4$ tail can be extracted with good precision as discussed in 
the experimental papers \cite{stewart2010,kuhnle2010,wild2012}.
If we assume that this subtraction of leading order can be done without severe increase of 
uncertainties in the data, then one would need to look at small and intermediate $k$ 
values for this sub-leading tail behavior.

A natural extension of the present work is to consider three-body states in dimensions 
lower than three. In two-dimensions it is well-known that no Efimov effect 
occurs \cite{bru79,adh88,nie97,nie01,artem2013} and, among other things, this 
implies that the momentum 
distribution does not have the subleading oscillatory behavior in two dimensions \cite{bel13a}.
For two-dimensional systems with large differences
in the masses it is still possible to have many three-body bound states \cite{bel13b} and 
this should also be reflected in some way through the asymptotic momentum distribution.
Another intriguing question is how the universal tail behavior and the contact
relations behave in a crossover between two- and three-dimensional or one- and three-dimensional 
setups \cite{pricoupenko2011,valiente2012,bel13a}.

While we have studied only short-range interactions in the current paper, it would be interesting to
consider the momentum tails of few-body bound states in systems with long-range interactions. 
Recent experiments with heteronuclear molecules have demonstrated that the momentum distribution 
in dipolar systems can be probed using absorption imaging \cite{wang2010}. Few-body
bound states of dipolar particles have been predicted in a large parameter regime for 
both one- \cite{klawunn2010a,zinner2011} and two-dimensional 
systems \cite{wang2006,klawunn2010b,baranov2011,zinner2012}. 
As was recently shown, one-dimensional dipolar few-body systems can in 
some cases be described by using zero-range interaction terms with appropriately chosen
effective interactions parameters \cite{vol2013}. This opens up the possibility of using the 
same formalism with short-range interactions as discussed in the current paper but applied in a
one-dimensional setup. It should then be possible to derive the contact parameters in 
the presence of few-body bound states with dipolar particles in one dimension, similarly
to what has been done for non-dipolar bosons \cite{olshanii2003} and fermions \cite{barth2011}.

\acknowledgments
We thank Robert Heck, Juris Ulmanis, and Rico Peres from the group of Matthias Weidem{\"u}ller in Heidelberg for updates
on the experimental progress of $^{6}$Li-$^{133}$Cs mixed systems, and F{\'e}lix Werner for discussions on the work
presented in Ref.~\cite{castin2011}. MTY and TF thanks the hospitality of the Department of Physics and Astronomy of the Aarhus University, where part of this work was done, and to the Brazilian agencies CNPq  and FAPESP. This work was supported in part by a grant
from the Danish Ministry of Science, Innovation, and Higher Education under the International Network program.

\appendix

\section{Derivation of the sub-leading terms}\label{appendix1} 

\subsection{$n_2$ term}
In equation \eqref{n2asym} we have a sub-leading term of the form
\begin{align}\label{sublead}
&\int d^3q_A \left|\chi_{AB}(q_A)\right|^2\left[
 \frac{2}{\left(q_A^2+\vec q_A\cdot \vec
q_B+q_B^2\;\frac{{\cal A}+1}{2{\cal A}}\right)^2}-\frac{8{\cal A}^2}{({\cal A}+1)^2}\frac{1}{q_B^4}\right]=&\nonumber\\
&\left|c_{AB}\right|^2 \int \frac{d^3q_A}{q_{A}^{4}} \left[
 \frac{1}{\left(q_A^2+\vec q_A\cdot \vec
q_B+q_B^2\;\frac{{\cal A}+1}{2{\cal A}}\right)^2}-\frac{4{\cal
A}^2}{({\cal A}+1)^2}\frac{1}{q_B^4}\right]= \frac{2\pi
\left|c_{AB}\right|^2}{q_{B}^{5}}\int
\frac{dx}{x^2} \left[
 \frac{1}{x^4+\frac{1}{{\cal A}}x^2+(\frac{{\cal A}+1}{2{\cal A}})^2}-\frac{1}{(\frac{{\cal A}+1}{2{\cal A}})^2}\right],&
\end{align}
where in the first equality we have inserted the asymptotic form of
$|\chi_{AB}(q_A)|^2=|c_{AB}|^3q_{A}^{-4}/2$ obtained after averaging over the oscillatory term 
in Eq.~\eqref{chiasymp}. In the
second equality we have performed the angular integral and
introduced the variable $q_A=q_B x$. We have also used the fact that
the integrand is even to extend the integration to the entire real
axis.

The function under the integral,
\begin{align}
f(x)=\frac{1}{x^2} \left[
 \frac{1}{x^4+\frac{1}{{\cal A}}x^2+(\frac{{\cal A}+1}{2{\cal A}})^2}-\frac{1}{(\frac{{\cal A}+1}{2{\cal A}})^2}\right],
\end{align}
falls off faster than $1/x$ for $|x|\to\infty$. We can therefore
extend it to the complex domain and consider a contour in the
upper-half plane (or lower-half) that includes the real axis and a
semi-circle of large radius in a counterclockwise orientation. To
use the residue theorem, we need to first find the poles of $f(x)$.
Since $f(x)$ is regular at $x=0$, the only poles are out in the
complex plane. The four poles are given by
\begin{align}
x_1=r e^{i\theta_1/2},\,\,x_2=r e^{i(\pi-\theta_1/2)},\,\,x_3=r
e^{i(\pi+\theta_1/2)},\,\,x_4=r e^{-i\theta_1/2},
\end{align}
where $r=\sqrt{\tfrac{{\cal A}+1}{2{\cal A}}}$ and
$\tan^2\theta_1={\cal A}({\cal A}+2)$. If we use the convention that
$\pi/2<\theta_1<\pi$ as in the main text, then $x_1$ and $x_2$ are
the poles in the upper-half plane. The sum of the two residues is
\begin{align}
\textrm{Res}(f,x_1)+\textrm{Res}(f,x_2)=-\frac{1}{ir^3}\frac{{\cal
A}({\cal A}+3)}{({\cal
A}+1)^2}\frac{\cos(\frac{\theta_1}{2})}{\sin(\theta_1)}
\end{align}
Using the residue theorem, the sub-leading term in Eq.~\eqref{sublead} then becomes
\begin{align}
&\int d^3q_A \left|\chi_{AB}(q_A)\right|^2\left[
 \frac{2}{\left(q_A^2+\vec q_A\cdot \vec
q_B+q_B^2\;\frac{{\cal A}+1}{2{\cal A}}\right)^2}-\frac{8{\cal
A}^2}{({\cal A}+1)^2}\frac{1}{q_B^4}\right]= -\frac{4\pi^2
\left|c_{AB}\right|^2}{q_{B}^{5} 2\sin(\frac{\theta_1}{2})}\frac{{\cal
A}({\cal A}+3)}{({\cal A}+1)^2} \left(\frac{2{\cal A}}{{\cal
A}+1}\right)^{3/2}.
\end{align}
From the definition of $\theta_1$ we see that $cos
\theta_1=-\frac{1}{{\cal A}+1}$ and 
$\left[2\sin(\frac{\theta_1}{2})\right]^{-1}=\sqrt{\frac{{\cal A}+1}{2({\cal A}+2)}}$. The sub-leading term in $n_2$ is given by
\begin{equation}
\left<n_2(q_B)\right>=-\frac{8\pi^2 \left|c_{AB}\right|^2}{q_{B}^{5}} \frac{{\cal
A}^3({\cal A}+3)}{({\cal A}+1)^3 \sqrt{{\cal A}({\cal A}+2)}} \ ,
\end{equation}
where the special case ${\cal A}=1$ yields $\theta_1=2\pi/3$ and
$\left<n_2\right>=-4\pi^2 \left|c_{AB}\right|^2/(\sqrt{3}q_{B}^{5})$.

\subsection{$n_3$ term}\label{appendix2}
Neglecting the three-body energy and making
the variable transformation $\vec{q}_A=\vec{p}_B-\frac{\vec{q}_B}{2}$ in equation \eqref{n3}, we find
\begin{eqnarray}
n_3(q_B)&=& 2\chi^{\ast}_{AA}(q_B)
  \int d^3q_A \frac{\chi_{AB}(
q_A)}{\left(q^{2}_A+{\vec q}_A\cdot{\vec{q}_B}+q_B^2\frac{{\cal
A}+1}{2{\cal A}}\right)^2}+c.c. \ .
\end{eqnarray}

Defining ${\vec q}_A=q_B{\vec y}$, integrating over the solid angle and replacing the asymptotic form for the spectator 
functions $\chi_{AA}$ and $\chi_{AB}$, given by equation (\ref{chiasymp}), we get 
\begin{eqnarray}
n_3(q_B)&=& 8 \pi \frac{c^{\ast}_{AA}\;c_{AB}}{q_B^5}\sin^2(s\;\log q_B/q^*)
  \int_0^\infty \frac{\cos(s\;\log y) \ dy }{y^{4}+\frac{1}{{\cal A}}y^{2}+\left(\frac{{\cal
A}+1}{2{\cal A}}\right)^2} \nonumber\\ 
&+& 8 \pi \frac{c^{\ast}_{AA}\;c_{AB}}{q_B^5}\sin(s\;\log q_B/q^*)\cos(s\;\log q_B/q^*)
  \int_0^\infty \frac{\sin(s\;\log y) \ dy }{y^{4}+\frac{1}{{\cal A}}y^{2}+\left(\frac{{\cal
A}+1}{2{\cal A}}\right)^2}+c.c. \ . \label{n3int}
\end{eqnarray}
Averaging out the oscillatory terms, only the first term of equation (\ref{n3int}) gives a 
non-vanishing result 
\begin{equation}
\langle n_3(q_B)\rangle= 4 \pi \frac{c^{\ast}_{AA}\;c_{AB}}{q_B^5}
  \int_0^\infty \frac{\cos(s\;\log y) \ dy }{y^{4}+\frac{1}{{\cal A}}y^{2}+\left(\frac{{\cal
A}+1}{2{\cal A}}\right)^2}+c.c. 
\label{n3avg}
\end{equation}
Expressing cosine in the complex exponential form we write that
\begin{eqnarray}
I&=&\int_0^\infty \frac{\cos(s\;\log y) \ dy }{y^{4}+\frac{1}{{\cal A}}y^{2}+\left(\frac{{\cal
A}+1}{2{\cal A}}\right)^2} =  \Re\left[ \int_0^\infty \frac{y^{\imath s} \ dy }{y^{4}+\frac{1}{{\cal A}}y^{2}+\left(\frac{{\cal A}+1}{2{\cal A}}\right)^2}\right]= \Re\;I_1,   
\label{intn3}
\end{eqnarray}
where $\Re$ denotes the real part. The residue theorem can be applied to solve the above integral. We set $y=e^\alpha$ in order to extend the interval of integration from $-\infty$ to $\infty$ and rewrite $I_1$ as
\begin{eqnarray}
I_1&=&\int_{-\infty}^\infty{ \frac{e^{\alpha(1+\imath s)} }{\left(e^\alpha-e^{\alpha_1}\right) \left(e^\alpha-e^{\alpha_2}\right) \left(e^\alpha-e^{\alpha_3}\right) \left(e^\alpha-e^{\alpha_4}\right)} d\alpha } \ .
\end{eqnarray}
The next steps are about extending the integrand, $f(\alpha)$, to the complex plan, finding its poles and evaluating the residues of the poles. All the roots in the denominator of $f(\alpha)$ are in the complex plane, out of the real axis and are given by
\begin{eqnarray}
\alpha_1=\ln r + \imath \theta_3 , \ \alpha_2=\ln r - \imath (\pi-\theta_3), \ \alpha_3=\ln r - \imath \theta_3, \ \alpha_4=\ln r + \imath (\pi-\theta_3) \ , 
\label{rootsn3}
\end{eqnarray}
with $r=\sqrt{\frac{{\cal A}+1}{2{\cal A}}}$ and $\tan\theta_3=\sqrt{\frac{{\cal A}+2}{{\cal A}}}$ for $0\leq\theta_3\leq\pi/2$. 

We extend $f(\alpha)$ to the complex plane and choose the closed path as a rectangle of vertices $-R$, $+R$, $+R+\imath\pi$ and $-R+\imath\pi$ (for $R\to\infty$), which encompasses the poles $\alpha_1$ and $\alpha_4$ in the upper-half plane. We are left with four integrals, namely $J_1$ which extends along the real axis from $-R$ to $+R$, $J_2$ from $+R$ to $+R+\imath\pi$, $J_3$ from $+R+\imath\pi$ to $-R+\imath\pi$ and $J_4$ from $-R+\imath\pi$ to $-R$. In the limit $R\to\infty$ we find that $J_1=I_1$, $J_3=e^{-s\pi}I_1$ and $J_3\;\text{and}\;J_4\to0$.  In this way we find that
\begin{equation}
I_1=\frac{2\pi\imath}{1+e^{-\pi s}}\left[Res(f,\alpha_1)+Res(f,\alpha_4)\right]=\frac{\pi {\cal A}} {1+e^{-\pi s}}\sqrt{\frac{2}{({\cal A}+2)({\cal A}+1)}}\left(e^{\imath s(\ln r+\imath\theta_3)-\imath\theta_3}+e^{\imath s(\ln r+\imath(\pi-\theta_3))+\imath\theta_3}\right) \ ,
\end{equation}
where $r$ and $\theta_3$ are defined after equation \eqref{rootsn3}.

It is necessary to split the real, $\Re$, and imaginary, $\Im$, parts to achieve our goal. Manipulating the trigonometric and hyperbolic functions we get
\begin{eqnarray}
\Re\;I_1&=&\frac{\pi}{2\cosh\left(\frac{s\pi}{2}\right)}\left\{\sqrt{\frac{{\cal A}}{{\cal A}+2}}\cos\left(s\ln r \right)\cosh\left[s\left(\frac{\pi}{2}-\theta_3\right)\right]+\sin\left(s\ln r \right)\sinh\left[s\left(\frac{\pi}{2}-\theta_3\right)\right]\right\} \ , \label{n3re} \\
\Im\;I_1&=&\frac{-\pi}{2\cosh\left(\frac{s\pi}{2}\right)}\left\{\sqrt{\frac{{\cal A}}{{\cal A}+2}}\sin\left(s\ln r \right)\cosh\left[s\left(\frac{\pi}{2}-\theta_3\right)\right]+\cos\left(s\ln r \right)\sinh\left[s\left(\frac{\pi}{2}-\theta_3\right)\right]\right\} \ . \label{n3im}
\end{eqnarray}

Finally, from equations \eqref{n3avg}, \eqref{intn3} and \eqref{n3re}, the non-oscillating part of $n_3(q_B)$ is given by
\begin{equation}
\langle n_3(q_B)\rangle= \frac{4 \pi^2 c_{AA}\;c_{AB}}{q_B^5 \cosh\left(\frac{s\pi}{2}\right)}\left\{\sqrt{\frac{{\cal A}}{{\cal A}+2}}\cos\left(s\ln \sqrt{\frac{{\cal A}+1}{2{\cal A}}}\right)\cosh\left[s\left(\frac{\pi}{2}-\theta_3\right)\right]+\sin\left(s\ln \sqrt{\frac{{\cal A}+1}{2{\cal A}}}\right)\sinh\left[s\left(\frac{\pi}{2}-\theta_3\right)\right]\right\} \ ,
\end{equation}
where $\tan\theta_3=\sqrt{\frac{{\cal A}+2}{{\cal A}}}$ for $0\leq\theta_3\leq\pi/2$. The special case ${\cal A}=1$ yields $\theta_3=\pi/3$ and $\langle n_3(q_B)\rangle= 4 \pi^2 |c_{AA}|^2 \cosh\left(\frac{s\pi}{6}\right) /\left(q_B^5 \sqrt{3} \cosh\left(\frac{s\pi}{2}\right)\right)$.

\subsection{$n_4$ term}
Although equations \eqref{n3} and \eqref{n4} are similar, it is not possible to extend the results in appendix \ref{appendix2} to obtain the non-oscillating term of $n_4(q_B)$. Defining $\vec{p}_B=\frac{\vec{q}_B}{2}\vec{y}$ and dropping the three-body energy, equation \eqref{n4} becomes
\begin{equation}
n_4(q_B)= \frac{4\pi}{q_B}\int_0^\infty{\frac{y^2 dy}{\left(y^2+\frac{{\cal A}+2}{{\cal A}}\right)^2}}\int_{-1}^{+1} dx\;
 \chi^{\ast}_{AB}(q_B x_-)\chi_{AB}(q_B x_+)+c.c. \ ,
\end{equation}
where $x_\pm=\frac{1}{2}\sqrt{1+y^2\pm2yx}$. Replacing the spectator function by its asymptotic form \eqref{chiasymp} in the integral above, we are left with three terms, which read
\begin{eqnarray}
n_4(q_B)&=& \frac{8\pi |c_{AB}|^2 \sin^2\left(s \ln\frac{q_B}{q^\ast}\right)}{q_B^5}\int_0^\infty{\frac{y^2 dy}{\left(y^2+\frac{{\cal A}+2}{{\cal A}}\right)^2}}\int_{-1}^{+1} \frac{dx}{x_+^2x_-^2}\cos\left(s \ln x_+\right)\cos\left(s \ln x_-\right) \nonumber\\
&+&\frac{8\pi |c_{AB}|^2 \cos^2\left(s \ln\frac{q_B}{q^\ast}\right)}{q_B^5}\int_0^\infty{\frac{y^2 dy}{\left(y^2+\frac{{\cal A}+2}{{\cal A}}\right)^2}}\int_{-1}^{+1} \frac{dx}{x_+^2x_-^2}\sin\left(s \ln x_+\right)\sin\left(s \ln x_-\right) \nonumber \\ 
&+&\frac{4\pi |c_{AB}|^2 \sin\left(s \ln\frac{q_B}{q^\ast}\right)\cos\left(s \ln\frac{q_B}{q^\ast}\right)}{q_B^5}\int_0^\infty{\frac{y^2 dy}{\left(y^2+\frac{{\cal A}+2}{{\cal A}}\right)^2}}\int_{-1}^{+1} \frac{dx}{x_+^2x_-^2}\sin\left[s\ln\left(x_+ x_-\right)\right] \ . \label{n4osc}
\end{eqnarray} 
As it was done for $n_3(q_B)$, averaging out the oscillatory term, only the two first terms on the right-hand-side of equation \eqref{n4osc} give a non-vanishing contribution. The angular integration is performed using that
\begin{equation}
\int{dx\left(\frac{\beta+x}{\beta-x}\right)^{\pm\imath s/2}\left(\beta^2-x^2\right)^{-1}}=\pm \left(\frac{\beta+x}{\beta-x}\right)^{\pm\imath s/2} \left(\imath \beta s\right)^{-1}
\end{equation}
and the non-oscillating part of $n_4(q_B)$ is given by
\begin{equation}
\left\langle n_4(q_B) \right\rangle=\frac{32\pi |c_{AB}|^2 }{\imath\;s\; q_B^5}\int_0^\infty{\frac{y\; dy}{\left(y^2+\frac{{\cal A}+2}{{\cal A}}\right)^2 (1+y^2)}\left[\left(\frac{y+1}{|y-1|}\right)^{\imath s}-\left(\frac{y+1}{|y-1|}\right)^{-\imath s}\right]} \ .
\label{n4avg}
\end{equation}
As was pointed out in \cite{castin2011}, the absolute value complicates the calculation of this integral. Circumventing this problem, we follow the same trick as in \cite{castin2011}, where the integral is split in two pieces: $y \in [0,1]$ and $y \in \left[\right.1,\infty\left[\right.$ and a new variable is introduced in each piece. We set $y=\frac{x-1}{x+1}$ in the first piece and $y=\frac{x+1}{x-1}$ in the second piece. Notice that in both cases $x \in [1,\infty[$. 
Now we are able to apply the residue theorem to calculate the non-oscillating part of $n_4(q_B)$. First of all we introduce a new variable $\alpha$, such that $x=e^\alpha$ and $\alpha \in [0,\infty[$. As the resulting integrand is even, it allows us to extend the domain of integration to the entire real axis, i.e. $\alpha \in\; ]-\infty,\infty[$. The integral in equation \eqref{n4avg} reads
\begin{eqnarray}
I&=& \frac{\imath\;{\cal A}^2}{ ({\cal A}+1)^4}\Im\left[\int_{-\infty}^\infty{\frac{e^{\alpha(1+\imath s)} \left(e^{2\alpha}-1\right) \left[({\cal A}+1)^2 \left(e^{6\alpha}+1\right)/2+(3{\cal A}^2-2{\cal A}-1) \left(e^{2\alpha}+e^{4\alpha} \right)/2\right]} {\left(e^\alpha-e^{\alpha_5}\right) \left(e^\alpha-e^{\alpha_6}\right)\left[ \left(e^\alpha-e^{\alpha_1}\right) \left(e^\alpha-e^{\alpha_2}\right) \left(e^\alpha-e^{\alpha_3}\right) \left(e^\alpha-e^{\alpha_4}\right)\right]^2}\;d\alpha}\right] \\
&=&\frac{\imath\;{\cal A}^2}{({\cal A}+1)^4}\Im\;I_1,
\label{n4int}
\end{eqnarray}
where $\Im$ denotes the imaginary value.
Extending the integrand, $f(\alpha)$, to the complex plane, we find that all the roots in its denominator are on the imaginary axis and given by
\begin{equation}
\alpha_1=\imath \theta_4, \ \alpha_2=\imath (\pi-\theta_4), \ \alpha_3=-\imath \theta_4, \ \alpha_4=-\imath (\pi-\theta_4), \ \alpha_5=\imath \frac{\pi}{2}, \ \alpha_6=-\imath \frac{\pi}{2} \ ,
\label{n4roots}
\end{equation} 
where $\tan\theta_4=\sqrt{{\cal A}({\cal A}+2)}$ for $0\leq\theta_4\leq\frac{\pi}{2}$. Notice that $\alpha_5$ and $\alpha_6$ are simple poles while $\alpha_1$, $\alpha_2$, $\alpha_3$, and $\alpha_4$ are poles of second order (see Eq.~(\eqref{n4int})).

To evaluate the contour integral, we choose the closed path as in the calculation of $n_3(q_B)$, namely a rectangle of vertices $-R$, $+R$, $+R+\imath\pi$ and $-R+\imath\pi$ (for $R\to\infty$), which encompasses the poles $\alpha_1$, $\alpha_2$ and $\alpha_5$ in the upper-half plane. Once more we are left with four integrals, i.e. $J_1$ which extends along the real axis from $-R$ to $+R$, $J_2$ from $+R$ to $+R+\imath\pi$, $J_3$ from $+R+\imath\pi$ to $-R+\imath\pi$ and $J_4$ from $-R+\imath\pi$ to $-R$. In the limit $R\to\infty$ we find that $J_1=I_1$, $J_3=e^{-s\pi}I_1$ and $J_3\;\text{and}\;J_4\to0$.  In this way we find that
\begin{equation}
I_1=\frac{2\pi\imath}{1+e^{-\pi s}}\left[Res(f,\alpha_1)+Res(f,\alpha_2)+Res(f,\alpha_5)\right] \ .
\end{equation}

Calculating the residues is tedious, except for the case of $\alpha_5$ where $Res(f,\alpha_5)=0$. After some algebraic work, the real and imaginary part of $I_1$ are given by
\begin{eqnarray}
\Re\;I_1&=& \frac{\pi ({\cal A}+1)^3 {\cal A}}{4 \sqrt{{\cal A}({\cal A}+2)} \cosh\left(\frac{s\pi}{2}\right)}\cosh\left[s\left(\frac{\pi}{2}- \theta_4\right) \right] \ , \\
\Im\;I_1&=&\frac{\pi ({\cal A}+1)^4}{4 \sqrt{{\cal A}({\cal A}+2)} \cosh\left(\frac{s\pi}{2}\right)} \left\{\sqrt{{\cal A}({\cal A}+2)}\sinh\left[s\left(\frac{\pi}{2}- \theta_4\right) \right]-\frac{s\; {\cal A} }{{\cal A}+1}\cosh\left[s\left(\frac{\pi}{2}- \theta_4\right) \right]\right\} \ . \label{n4im}
\end{eqnarray} 

Combining Eqs.~\eqref{n4avg}, \eqref{n4int}, and \eqref{n4im}, the non-oscillating part of $n_4(q_B)$ finally reads
\begin{equation}
\left\langle n_4(q_B) \right\rangle=\frac{8\pi^2 |c_{AB}|^2 {\cal A}^2 }{s\; q_B^5\; \cosh\left(\frac{s\pi}{2}\right)}\left\{ \sinh\left[s\left(\frac{\pi}{2}- \theta_4\right) \right]-\frac{s\; {\cal A} }{\sqrt{{\cal A}({\cal A}+2)} ({\cal A}+1)}\cosh\left[s\left(\frac{\pi}{2}- \theta_4\right) \right]\right\} \ ,
\end{equation} 
where $\tan\theta_4=\sqrt{{\cal A}({\cal A}+2)}$ for $0\leq\theta_4\leq\pi/2$. The special case ${\cal A}=1$ yields $\theta_4=\pi/3$ and $\langle n_4(q_B)\rangle= 8 \pi^2 |c_{AA}|^2\left[\sinh\left(\frac{s\pi}{6}\right)-s/(2\sqrt{3})\cosh\left(\frac{s\pi}{6}\right) \right]  /\left[s\; q_B^5\;  \cosh\left(\frac{s\pi}{2}\right)\right]$.

\end{document}